\documentclass[nofootinbib,a4paper]{revtex4}

\usepackage{graphicx}%
\usepackage{multirow}%
\usepackage{amsmath,amssymb,amsfonts}%
\usepackage{amsthm}%
\usepackage{mathrsfs}%
\usepackage[title]{appendix}%
\usepackage{xcolor}%
\usepackage{textcomp}%
\usepackage{booktabs}%
\usepackage{algorithm}%
\usepackage{algorithmicx}%
\usepackage{algpseudocode}%
\usepackage{listings}%
\usepackage{verbatim}
\usepackage{xspace}
\usepackage{orcidlink}

\DeclareSymbolFontAlphabet{\mathrsfs}{rsfs}
\newcommand{\scri}{\mathrsfs{I}}
\newcommand{\scripx}{$\scri^+$}
\newcommand{\scrip}{\scripx\xspace}

\newcommand{\K}{K_{\textrm{CMC}}}
\newcommand{\aconf}{\bar\Omega}
\newcommand{\rscri}{r_{\!\!\scri}}
\newcommand{\backbetar}{\hat{\beta^r}}
\newcommand{\backalpha}{\hat{\alpha}}
\newcommand{\nconst}{n_{cK}}
\newcommand{\cnK}{\xi_{cK}}
\newcommand{\rmatch}{\tilde r_{\textrm{match}}}
\newcommand{\rcompmatch}{r_{\textrm{match}}}

\newcommand{\eref}[1]{(\ref{#1})}
\newcommand{\sref}[1]{section~\ref{#1}}
\newcommand{\Sref}[1]{Section~\ref{#1}}
\newcommand{\ssref}[1]{subsection~\ref{#1}}
\newcommand{\aref}[1]{appendix~\ref{#1}}
\newcommand{\fref}[1]{figure~\ref{#1}}
\newcommand{\Fref}[1]{Figure~\ref{#1}}

\newcommand{\CZ}{Z4c}
\newcommand{\thesis}{Vano-Vinuales:2015lhj}
\newcommand{\procere}{Vano-Vinuales:2014ada}
\newcommand{\procmg}{Vano-Vinuales:2016mbo}
\newcommand{\pap}{Vano-Vinuales:2014koa}
\newcommand{\papgauge}{Vano-Vinuales:2017qij}
\newcommand{\papbh}{Vano-Vinuales:2023yzs}
\newcommand{\pappen}{Vano-Vinuales:2023pum}

\begin{document}

\title{Height-function-based 4D reference metrics for hyperboloidal evolution}

\author{Alex Va\~n\'o-Vi\~nuales\,\orcidlink{0000-0002-8589-006X}}   \author{Tiago Valente\,\orcidlink{0009-0003-4713-2561}}   \address{Centro de Astrof\'{\i}sica e Gravita\c c\~ao - CENTRA, Departamento de F\'isica, Instituto Superior T\'ecnico IST, Universidade de Lisboa UL, Avenida Rovisco Pais 1, 1049-001 Lisboa, Portugal}

\begin{abstract}
Hyperboloidal slices are spacelike slices that reach future null infinity. Their asymptotic behaviour is different from Cauchy slices, which are traditionally used in numerical relativity simulations. 
This work uses free evolution of the formally-singular conformally compactified Einstein equations in spherical symmetry. One way to construct gauge conditions suitable for this approach relies on building the gauge source functions from a time-independent background spacetime metric. This background reference metric is set using the height function approach to provide the correct asymptotics of hyperboloidal slices of Minkowski spacetime. 
The present objective is to study the effect of different choices of height function on hyperboloidal evolutions via the reference metrics used in the gauge conditions. A total of 10 reference metrics for Minkowski are explored, identifying some of their desired features. They include 3 hyperboloidal layer constructions, evolved with the non-linear Einstein equations for the first time. Focus is put on long-term numerical stability of the evolutions, including small initial gauge perturbations. The results will be relevant for future (puncture-type) hyperboloidal evolutions, 3D simulations and the development of coinciding Cauchy and hyperboloidal data, among other applications. 
\end{abstract}


\maketitle

\section{Introduction}\label{intro}

Numerical simulations of gravitational-wave-emitting astrophysical sources, such as compact binaries of black holes or neutron stars, can be modeled as isolated systems. Of interest is not only the central part, where the physics of the system is to be described as accurately as possible, but also the far-field region where the gravitational waves propagate. Their signals can only be unambiguously defined at future null infinity (\scrip), the collection of end-points of future-directed null geodesics. A possible way of solving the Einstein equations as an initial value formulation \textit{including future null infinity} is via evolution on hyperboloidal slices \cite{friedrich1983,lrr-2004-1,Friedrich:2003fq}. They are spacelike everywhere, but become asymptotically tangent to characteristic ones and so are able to reach \scrip. 
This work is part of a research line \cite{\pap,\procere,\procmg,\papgauge,\papbh} motivated by \cite{Zenginoglu:2007it,Zenginoglu:2008pw} that solves the Einstein equations on hyperboloidal slices using free evolution, and expands on the construction of gauge conditions in spherical symmetry. The objective is to create the infrastructure to perform 3D hyperboloidal simulations for the puncture approach \cite{Brandt:1997tf,Campanelli:2005dd,Baker:2005vv}. Work towards an alternative method to free hyperboloidal evolution with the dual frame approach \cite{Hilditch:2015qea} is currently ongoing \cite{Peterson:2024bxk}, in addition to other efforts \cite{Bardeen:2011ip,Rinne:2013qc,Frauendiener:2023ltp}. 

A critical difference between evolution with traditional Cauchy slices and hyperboloidal ones is the asymptotic behaviour of the fields. For instance, asymptotically on a Cauchy slice the radial component of the shift $\beta^{\tilde r}\to 0$ and the trace of the extrinsic curvature $\tilde K\to 0$. This is not the case on a hyperboloidal slice, where those quantities asymptote to a non-vanishing value. Gauge conditions employed in Cauchy-based simulations, such as harmonic (see e.g.~\cite{Friedrich:2000qv}), the Bona-Mass\'o family of slicing conditions \cite{Bona:1994dr}, or the Gamma-driver shift condition \cite{Alcubierre:2002kk}, thus require adaptations for use with the hyperboloidal method. 
One possible approach \cite{\papgauge} is to construct gauge source functions from a time-independent background metric adapted to the asymptotic behaviour of the hyperboloidal slice, like a ``hyperboloidal'' reference metric. 

The reference metric approach to evolutions of the Einstein equations, used in earlier works like \cite{Bonazzola:2003dm,Garfinkle:2007yt}, crystallized with the (covariant BSSN) formulation by Brown \cite{Brown:2007nt,Brown:2009dd}, which is well suited for the use of curvilinear coordinate systems. To appreciate its advantages, note that the traditional BSSN formalism \cite{NOK,PhysRevD.52.5428,Baumgarte:1998te} has the contracted spatial Christoffel symbol $\Gamma^a=\gamma^{bc}\Gamma^a_{bc}$ as evolution variable. However, this choice has two main disadvantages: the transformations between BSSN variables include tensor weights and, if expressed in spherical coordinates, $\Gamma^r$ takes a singular value at the origin even for Minkowski. Brown's \cite{Brown:2009dd} formulation introduces $\Lambda^a$ (see \eref{Lambdadef}), which is constructed as a contracted difference of Christoffel symbols and as such transforms as a tensor. When expressing the equations in curvilinear coordinates, the subtracted connection $\hat \Gamma^a_{bc}$ takes away the divergent behaviour and $\Lambda^r$ is finite. 
The reference metric method has been used with spherical polar coordinates coordinates, together with a partially implicit time integrator, in \cite{Baumgarte:2012xy,Montero:2012yr,Sanchis-Gual:2014nha,Cordero-Carrion:2014zza}, 
and has also been extended to general relativistic hydrodynamics in \cite{Montero:2013pca}.  
The Z4 \cite{bona-2003-67,Alic:2011gg} equations include constraint propagation and damping terms, which have played an important role in the results of this work. The variations as described in \cite{Cao:2011fu,Sanchis-Gual:2014nha} and in the formulation used here, include terms with the reference metric similarly as in Brown's formulation. 
An interesting related construction, although used for different purposes, is the gauge-fixing DeTurck trick \cite{10.4310/jdg/1214509286,Headrick:2009pv,Dias:2015pda}, which turns the Ricci-flow equation strictly parabolic, and renders the Einstein equations strictly elliptic. It generalises the usual harmonic coordinate gauge in a way similar to the hyperbolic gauge conditions in section 4 of \cite{\papgauge}. In the DeTurck construction, a spacetime vector $\xi^a=g^{bc}\left(\Gamma^a_{bc}-\hat\Gamma^a_{bc}\right)$\footnote{Note that these are not spatial, but spacetime quantities, and $\hat\Gamma^a_{bc}$ is constructed from the given reference metric $\hat g_{ab}$.} is introduced and its symmetrized covariant derivative added to the Einstein equations, similarly as in the constraint addition of the Z4 and generalised harmonic formalisms. 

The introduction of a hyperboloidal 4D reference metric for the construction of the gauge conditions (explained in \sref{gauge}) appears naturally and provides a systematic way of setting the gauge source functions. There are infinitely many choices of hyperboloidal slices for the reference metric, and these may have a different behaviour in the simulations despite using the same functional form for the gauge conditions (which can also be modified).
An interesting setup are hyperboloidal layers. They consist of a purely Cauchy slice in the interior joined at some value of the radius to an outer hyperboloidal one, where the latter is referred to as the layer. This construction was first introduced in \cite{Zenginoglu:2010cq}, and then applied mainly to self-force calculations of black hole perturbation theory and modelling in the extreme-mass-ratio limit, such as \cite{Bernuzzi:2010xj,Bernuzzi:2011aj,Zenginoglu:2011zz,Bernuzzi:2012ku,Harms:2013ib,Harms:2014dqa}. Hyperboloidal layers for Minkowski are set up here as reference metrics and evolved with the Einstein equations for the first time. 
This work thus explores different options for hyperboloidal reference metrics in spherical symmetry as a first step towards understanding the available options and their ability to provide long-term stable hyperboloidal evolutions. 

This paper is organised as follows: \sref{form} summarises the formulation of the Einstein equations used and \sref{gauge} gives the form of the gauge conditions. \Sref{refmet} explains the construction of the reference metrics using the height function approach. The considered options are described in \sref{refmetoptions}, with visual representations of the metrics included in \aref{pics}. The results of evolving the Einstein equations with the presented 4D hyperboloidal reference metrics are presented in \sref{resu}, and conclusions are drawn in \sref{concl}.

Notation and conventions: The chosen metric signature is $(-,+,+,+)$ and fundamental constants $G$ and $c$ are set to unity. The convention for the sign of the extrinsic curvature is that of \cite{Misner1973}, so that a negative value (e.g. asymptotically towards \scrip on a hyperboloidal slice) means expansion of the normals. The metric notation is the same as in \cite{\pap,\papbh}: the 4-dimensional physical spacetime metric is denoted as $\tilde g_{ab}$, the 4D conformal metric as $\bar g_{ab}$, the 3D conformal spatial metric (induced by $\bar g_{ab}$) as $\bar \gamma_{ab}$ and the 3D twice conformal metric as $\gamma_{ab}$. Hatted quantities are time-independent background quantities corresponding to or constructed from the chosen reference metric, such as the 3D twice conformal background metric $\hat \gamma_{ab}$ or the trace of the background physical extrinsic curvature $\hat{\tilde K}$.

\section{Formulation}\label{form}

Conformal compactification \cite{PhysRevLett.10.66} is a regularisation technique suitable when using compressed coordinates to reach infinity. The latter involves rescaling the radial coordinate to relocate infinity to a finite value of the coordinate radius. 
The (now diverging) physical metric $\tilde g_{ab}$ is expressed in terms of a rescaled metric $\bar g_{ab}$ that is finite via
\begin{equation}\label{rescmetric}
\bar g_{ab} = \Omega^2\tilde g_{ab} , 
\end{equation}
where the conformal factor satisfies $\Omega|_\scri=0$ and $\bar\nabla_a\Omega|_\scri\neq 0$.\footnote{These expressions hold at future null infinity (\scrip) and past null infinity ($\scri^-$), which have been jointly labeled as $\scri$. The same notation will be used in the rest of the text.}
Expressed in terms of the rescaled metric $\bar g_{ab}$, the 4D Einstein equations take the form
\begin{equation}\label{eq:EEconformal}
G_{ab}[\bar g] = 8\pi\ T_{ab} -\frac{2}{\Omega}\left(\bar \nabla_a\bar \nabla_b\Omega-\bar g_{ab}\bar \nabla^c\bar \nabla_c\Omega\right)-\frac{3}{\Omega^2}\bar g_{ab}(\bar \nabla_c\Omega)\bar \nabla^c\Omega . 
\end{equation}
The well-posed formulations considered are either the generalized/covariant \cite{Brown:2007nt,Brown:2009dd} BSSN-OK \cite{NOK,PhysRevD.52.5428,Baumgarte:1998te} equations and a conformal version of the Z4 \cite{bona-2003-67,Alic:2011gg,Sanchis-Gual:2014nha} formalism, the \CZ{} system \cite{Bernuzzi:2009ex,Weyhausen:2011cg}. The actual equations used in this work are included in appendix C of \cite{\pap} (or appendix A in \cite{\papgauge}) and in Chapter 2 of \cite{\thesis}, together with their complete derivation. 
The evolution variables are the 3D conformally rescaled spatial metric 
\begin{equation}
\gamma_{ab}=\chi\bar \gamma_{ab} , 
\end{equation}
where $\bar \gamma_{ab}$ is the spatial metric induced from $\bar g_{ab}$, and $\chi$ is the spatial conformal factor. The twice conformal spatial metric $\gamma_{ab}$ is assumed to have unit determinant. 
The evolved gauge variables are the conformal lapse $\alpha$ and the shift $\beta^i$. 
The conformal extrinsic curvature tensor $\bar K_{ab}$ is decomposed into its conformal trace-free part 
\begin{equation}
A_{ab}=\chi\bar K_{ab}-\frac{1}{3}\gamma_{ab}\bar K ,  \quad \textrm{with} \quad \bar K=\bar K_{ab}\bar\gamma^{ab}\equiv K_{ab}\gamma^{ab}, 
\end{equation}
 and (in this formulation) its physical trace, mixed with the physical (in the sense of non-conformally-rescaled) Z4 variable $\tilde \Theta$, 
\begin{equation}
\tilde K = \Omega\bar K-\frac{3\beta^a\partial_a\Omega}{\alpha}-2\tilde\Theta .  
\end{equation} 
Evolved are $A_{ab}$, and $\tilde K$'s variation $\Delta\tilde{K}=\tilde K-\hat{\tilde K}$ with respect to its background value $\hat{\tilde K}$, where the latter is calculated from the reference metric $\hat{\bar{g}}_{ab}$. The quantity $\tilde \Theta$ is evolved as well if using the Z4 formulation.
The Z4 variable $Z_a$ is absorbed into the vector 
\begin{equation}\label{Lambdadef}
\Lambda^a=\gamma^{bc}\left(\Gamma^a_{bc}-\hat\Gamma^a_{bc}\right) +2\gamma^{ab}Z_b ,
\end{equation} 
where $\Gamma^a_{bc}$ are the spatial Christoffel symbols calculated from the evolved $\gamma_{ab}$, and $\hat\Gamma^a_{bc}$ the ones built from the time-independent reference metric $\hat \gamma_{ab}$. 

The following ansatz is used for the spherically symmetric line element in the conformally compactified domain (with $d\sigma^2\equiv d\theta^2+\sin^2\theta d\phi^2$)
\begin{equation}\label{e:linel}
ds^2 = - \left(\alpha^2-\chi^{-1}\gamma_{rr}{\beta^r}^2\right) dt^2 + \chi^{-1}\left[2\, \gamma_{rr}\beta^r dt\,dr +  \gamma_{rr}\, dr^2 +  \gamma_{\theta\theta}\, r^2\, d\sigma^2\right] ,
\end{equation}
where the freedom introduced by the spatial conformal factor $\chi$ is fixed by imposing that the determinant of $\gamma_{ab}$ is one. This condition is used to eliminate $\gamma_{\theta\theta}$ in terms of $\gamma_{rr}$ as
\begin{equation}\label{delgtt}
\gamma_{\theta\theta}=\gamma_{rr}^{-1/2} . 
\end{equation}
In spherical symmetry, the only independent component of the trace-free part of the conformal extrinsic curvature $A_{ab}$ after explicitly imposing its trace-freeness is $A_{rr}$. Also only the radial component of the quantities $\Lambda^a$, $\beta^a$, $Z_a$ and the momentum constraint $M_a$ (denoted by $\Lambda^r$, $\beta^r$, $Z_r$ and $M_r$ respectively) remains non-zero. The evolution variables of the spherically symmetric reduced system are therefore $\chi$, $\gamma_{rr}$, $A_{rr}$, $\Delta\tilde K$, $\Lambda^r$, $\alpha$, $\beta^r$ and $\tilde\Theta$. They are to satisfy the Hamiltonian $H$, momentum $M_r$ and $Z_r$ constraints. 

\section{Construction of gauge conditions}\label{gauge}

The following gauge conditions have been adapted to the hyperboloidal setup as described in \cite{Vano-Vinuales:2017qij}. The principal part of the equations, which determines the hyperbolicity of the system, remains the same. The gauge source function terms, which may depend on the coordinates and metric components (but not on their derivatives) are chosen to have the same form as the rest of the equation. However the evolution quantities there are substituted by the equivalent ones constructed from the time-independent 4D reference metric. This structure becomes apparent in most terms in \eref{slicing} and \eref{shift}. The logic behind this setup is that when the evolution quantities are the same as the reference ones, as is expected in the stationary regime, the right-hand-sides of the gauge condition equations will be zero. 

The following, Bona-Mass\'o-inspired slicing condition has provided successful long-term hyperboloidal evolutions in the present spherically symmetric approach, with $\dot{} \equiv \partial_t$ and $' \equiv \partial_r$ , 
\begin{equation}\label{slicing} 
\dot\alpha = \beta^r\alpha' -\backbetar\backalpha'-\frac{(\nconst(\rscri^2-r^2)^2+\alpha^2)}{\Omega}\Delta\tilde K+\frac{\Omega'}{\Omega}(\backbetar\backalpha-\beta^r\alpha) + \frac{\cnK(\backalpha^2-\alpha^2)}{\Omega}  , 
\end{equation}
where $\cnK$ is a parameter used to damp the behaviour of the lapse at \scrip. This equation is equivalent to (20) in \cite{\papgauge} with $\xi_1=0$, $\xi_2=\cnK$ and $\alpha^2f= \nconst(\rscri^2-r^2)^2+\alpha^2$. 
The coefficient in front of $\Delta\tilde K$ is similar to the shock-avoiding slicing condition \cite{Alcubierre:1996su,Baumgarte:2022ecu,Li:2023pme}. 
This evolution equation for the lapse requires that the stationary value of $\Delta\tilde K$ be zero, as is the case for the setups considered in this work. 

The shift condition considered is a variant of the integrated Gamma-driver \cite{Alcubierre:2002kk} adapted to hyperboloidal slices 
\begin{equation}\label{shift}
\dot{{\beta^r}} = {\beta^r} {\beta^r}'-\backbetar \backbetar'+\left(\lambda (\rscri^2-r^2)^2 +\frac{3}{4}\alpha^2\chi\right) \Lambda^r+\eta  (\backbetar-{\beta^r}) . 
\end{equation}
This is the same as (26) in \cite{\papgauge} with $\xi_{\beta^r}=0$. The coefficient in front of $\Lambda^r$ is chosen in such a way that the associated propagation speeds will be the physical ones at \scrip. The positive parameter $\lambda$ increases the speeds in the interior of the domain, in an equivalent form to $\nconst$ for the slicing condition. For the effect of $\nconst$ and $\lambda$ see figure~1 in~\cite{\papgauge} for hyperboloidal slices of Minkowski and figure~1 in~\cite{\papbh} for those of Schwarzschild. 

\section{Construction of reference metrics}\label{refmet}

The choices for 4D spacetime reference metrics considered in this work correspond to Minkowski spacetime foliated along different hyperboloidal slices. The latter are characterized by the choice of height function $h(\tilde r)$ \cite{10.1063/1.524975,Gentle:2000aq,Malec:2003dq,Calabrese:2005rs,Zenginoglu:2007jw} employed in its construction. The height function relates the usual time coordinate $\tilde t$ to the hyperboloidal time coordinate $t$, whose level sets provide a hyperboloidal foliation, via 
\begin{equation}\label{trafot}
t = \tilde t-h(\tilde r) .
\end{equation}
Asymptotically, the height function must behave like the uncompactified radial coordinate $\tilde r$ ($-\tilde r$), which ensures that the hyperboloidal time asymptotes to the retarded (advanced) time and so reaches \scrip ($\scri^-$, past null infinity). Equivalently, the height function must satisfy that $|dh/d\tilde r| < 1$ everywhere except asymptotically, where $|dh/d\tilde r|_\scri=1$ holds, thus characterizing the hyperboloidal slices as spacelike but extending to $\scri$. The derivative $dh/d\tilde r$ for the height functions considered is shown in \fref{boostsum}. 

To reach future null infinity with a finite value of the spatial coordinates, the radial coordinate $\tilde r$ on a hyperboloidal slice is compressed into a new $r$ using a compactification factor $\aconf(r)$ 
\begin{equation}\label{trafor}
\tilde r=\frac{r}{\aconf(r)} .
\end{equation}
Following \eref{rescmetric}, the line element is conformally rescaled by the conformal factor $\Omega$, to provide regular metric components all the way to \scrip
\begin{equation}\label{confresc}
ds^2 = \Omega^2d\tilde s^2 .
\end{equation}
The compactification factor $\aconf$ is not to be confused with the conformal factor $\Omega$, as they are a priori different quantities and can be chosen to be different (see e.g. the choice of $\aconf$ for a hyperboloidal black hole trumpet slice in \cite{\papbh}). However, the conformal compactification method requires that both have the same (or at least proportional) behaviour near \scrip. 
Performing the previous transformations on the Minkowski line element ($d\tilde s^2 = -d\tilde t^2+d\tilde r^2+\tilde r^2 d\sigma^2$) yields
\begin{equation}\label{}
ds^2= -\Omega^2dt^2+\frac{\Omega^2}{\aconf^2}\left[-2\left(\partial_{\tilde r}h\right)(\aconf-r\,\aconf')dt\,dr+\left(1-\left(\partial_{\tilde r}h\right)^2\right)\frac{(\aconf-r\,\aconf')^2}{\aconf^2}d r^2 + r^2 d\sigma^2\right] ,
\end{equation}
Identification of the different parts of the metric \eref{e:linel} in the above line element after substituting \eref{delgtt} gives
\begin{subequations}\label{refmetcomp}
\begin{eqnarray}
\hat \chi{} = \frac{\aconf^2}{\Omega^2}\left(\frac{\aconf^2}{\left(1-\left(\partial_{\tilde r}h\right)^2\right)(\aconf-r\,\aconf')^2}\right)^{1/3} , &&
{\hat\gamma_{rr}} = \left(\frac{\left(1-\left(\partial_{\tilde r}h\right)^2\right)(\aconf-r\,\aconf')^2}{\aconf^2}\right)^{2/3} , \label{refmetcomp1} \\
{\hat\beta^r}= - \frac{\aconf^2\left(\partial_{\tilde r}h\right)}{\left(1-\left(\partial_{\tilde r}h\right)^2\right)(\aconf-r\,\aconf')}, &&
\hat\alpha= \Omega\sqrt{1-\left(\partial_{\tilde r}h\right)^2}. \label{refmetcomp2}
\end{eqnarray}
\end{subequations}
The reference value of the physical trace of the extrinsic curvature $\hat{\tilde K}$ and the initial value of its conformally rescaled component $A_{rr}{}_0$ are obtained substituting the above expressions into
\begin{subequations} \label{deriv}
\begin{eqnarray}
\hat{\tilde K} &=& \frac{\Omega \hat\beta^r{}'}{\hat{\alpha }}-\frac{3 \hat\beta^r \Omega \hat\chi '}{2 \hat{\alpha } \hat{\chi }}-\frac{3 \hat\beta^r \Omega'}{\hat{\alpha }}+\frac{\hat\beta^r \Omega \hat\gamma_{\theta\theta}'}{\hat{\alpha }   \hat\gamma_{\theta\theta}}+\frac{\hat\beta^r \Omega \hat\gamma_{rr}'}{2 \hat{\alpha } \hat\gamma_{rr}}+\frac{2 \hat\beta^r \Omega}{\hat{\alpha } r} , \\
A_{rr}{}_0 &=& \frac{{\hat\beta^r} {\hat\gamma_{rr}}'}{3 \hat\alpha }+\frac{2 {\hat\gamma_{rr}} {\hat\beta^r}{}'}{3 \hat\alpha }-\frac{{\hat\beta^r} {\hat\gamma_{rr}} {\hat\gamma_{\theta\theta}}'}{3 \hat\alpha  {\hat\gamma_{\theta\theta}}}-\frac{2 {\hat\beta^r} {\hat\gamma_{rr}}}{3 \hat\alpha  r} ,  \label{Arrini} 
\end{eqnarray}
\end{subequations}
where $\hat\gamma_{\theta\theta}$ is to be substituted using the hatted version of \eref{delgtt}.
Hatted quantities are included in the evolution equations as time-independent background data, and the initial values of the evolution variables are taken to be the same as the reference ones. Besides $A_{rr}{}_0$ above, we also have that $\Lambda^r{}_0=0$ (consequence of choosing $\gamma_{rr}{}_0=\hat\gamma_{rr}$), $\Delta\tilde K{}_0=0$ and $\tilde \Theta=0$. 

The reference metric (and the initial data) are fixed by choosing the height function $h$ in \eref{trafot}, the compactification factor $\aconf$ in \eref{trafor} and the conformal factor $\Omega$ in \eref{confresc}. The height function choices considered are described in \sref{refmetoptions}. A convenient further requirement is imposing that the twice conformal spatial reference metric be explicitly conformally flat, namely $\hat\gamma_{rr}=1$. This yields an ordinary differential equation for $\aconf$
\begin{equation}\label{aconfode}
\aconf'= \frac{\aconf}{r} \left(1-\frac{1}{\sqrt{1-\left(\partial_{\tilde r}h\right)^2}}\right) ,
\end{equation}
where the sign option providing the relevant solution was been chosen, and $\partial_{\tilde r}h$ is given in terms of the compactified $r$. Requiring \eref{aconfode} (and thus $\hat\gamma_{rr}=1$) is however not possible for all height functions considered (see examples in subsections \ref{cplusss} and \ref{cossinss}). 
Finally, the choice $\Omega=\aconf$ has been made in this work. This need not be the case, provided that $\Omega|_\scri\propto\aconf|_\scri$, so there is still more freedom to explore here, left for future work. 

\section{4D reference metrics}\label{refmetoptions}

Constant-mean-curvature (CMC) slices as described in \sref{cmcss} have been used in past work \cite{\pap,\procere,\procmg,\papgauge,\papbh,\pappen} in this research line, and are considered as the control case. The height functions from subsections \ref{cplusss} and \ref{cossinss} have a mixing of slice and compactification choices for which $\hat\gamma_{rr}=1$ is not possible. The reference metrics in subsections \ref{simss} and \ref{modss} are constructed from modifications of the CMC height function and can satisfy $\hat\gamma_{rr}=1$. Lastly, subsections \ref{layalmss} and \ref{laymodss} consider hyperboloidal layers: a setup with an inner Cauchy slice matched to a hyperboloidal one in the outer asymptotic region. 
A summary of the different choices with visualization of the most distinct ones in the form of conformal Carter-Penrose diagrams follows in \ssref{visu}. 

\subsection{Constant-mean-curvature height function -- labeled as ``CMC''}\label{cmcss}

The height function for CMC slices of Minkowski spacetime \cite{10.1063/1.524975,Zenginoglu:2007jw,Zenginoglu:2024bzs} (see e.g. \cite{brill1980k,Beig:1997fp,Gentle:2000aq} or \cite{\thesis,\papbh} for the derivation) takes the following simple form, expressed in terms of the uncompactified radial coordinate $\tilde r$ on the left and in the compactified radius $r$ on the right: 
\begin{equation} \label{cmch}
h(\tilde r) = \sqrt{\left(\frac{3}{\K}\right)^2+\tilde r^2}+\frac{3}{\K}\quad\longleftrightarrow\quad h(r) = \sqrt{\left(\frac{3}{\K}\right)^2+\left(\frac{r}{\aconf}\right)^2}+\frac{3}{\K} . 
\end{equation} 
The negative parameter $\K$ corresponds to the constant value of the trace of the extrinsic curvature $\hat{\tilde K} = \K$. The $\frac{3}{\K}$ term has been included so that the height function is zero for $\tilde r = r = 0$. This is not relevant for the construction for the reference metric, as the latter uses the radial derivative of the height function, which is called the boost. However, it is convenient when creating the conformal diagrams in \ssref{visu}, which do use $h$. 

Substitution of $\partial_{\tilde r}h$ from \eref{cmch} into the components of the reference metric \eref{refmetcomp} and use of \eref{trafor} gives 
\begin{subequations}
\begin{eqnarray}
\hat\chi{} = \frac{\aconf^2}{\Omega^2}\left(\frac{\left[\aconf^2+\left(\frac{\K\,r}{3}\right)^2\right]}{(\aconf-r\,\aconf')^2}\right)^{1/3}  , &&
\hat\gamma_{rr} = \left(\frac{(\aconf-r\,\aconf')^2}{\left[\aconf^2+\left(\frac{\K\,r}{3}\right)^2\right]}\right)^{2/3} , \\
{\hat\beta^r}=\frac{\left(\frac{\K\,r}{3}\right)\sqrt{\aconf^2+\left(\frac{\K\,r}{3}\right)^2}}{(\aconf-r\,\aconf')}, &&
\hat\alpha= \Omega\sqrt{1+\left(\frac{\K\,r}{3\aconf}\right)^2} . 
\end{eqnarray}
\end{subequations}
Solving \eref{aconfode} with the same substitution from \eref{cmch} yields the known solution
\begin{equation}\label{cmcaconf}
\aconf = \left(-\K\right)\frac{\rscri^2-r^2}{6\, \rscri}, 
\end{equation}
with $\rscri$ the coordinate location of future null infinity (set to $\rscri=1$ without restricting generality).
Setting $\Omega=\aconf$ satisfies that $\Omega$ is a regular function that becomes zero at $\scri$, with non-vanishing derivative there (compare e.g. \cite{Husa:2002zc,Schneemann}). It also provides the following simple expressions for the components of the 4D reference metric 
\begin{equation}\label{cmchatvals}
\hat\chi = \hat\gamma_{rr} = \hat\gamma_{\theta\theta} = 1 , \quad \hat\alpha = \sqrt{\Omega^2+\left(\frac{\K\,r}{3}\right)^2} \quad \textrm{and} \quad \hat\beta^r = \frac{\K\,r}{3}. 
\end{equation}

\begin{figure}[h]
\centering
\includegraphics[width=0.32\linewidth]{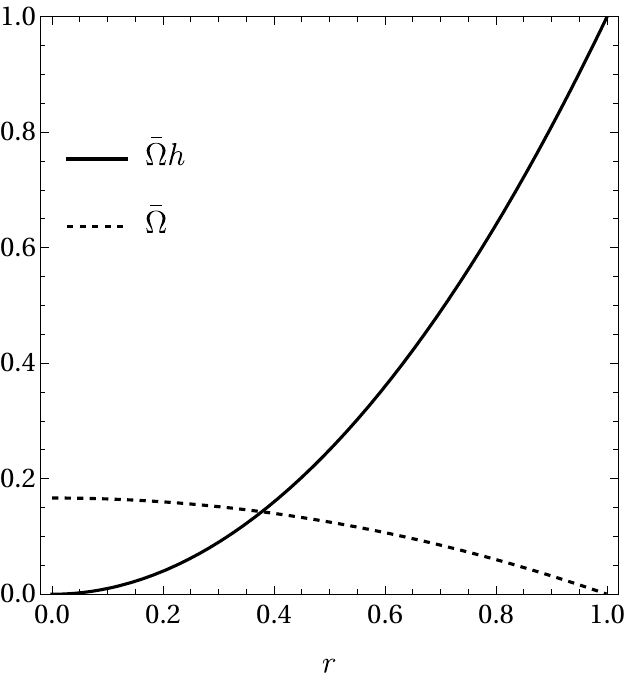}
\includegraphics[width=0.32\linewidth]{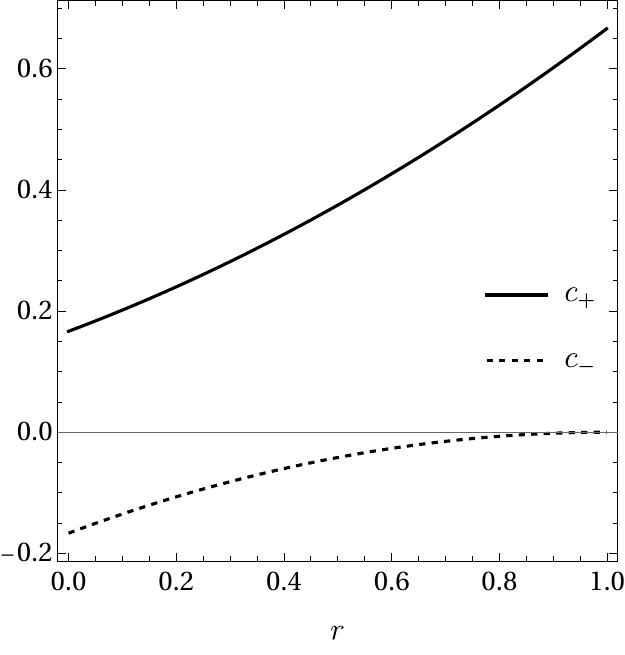}
\includegraphics[width=0.32\textwidth]{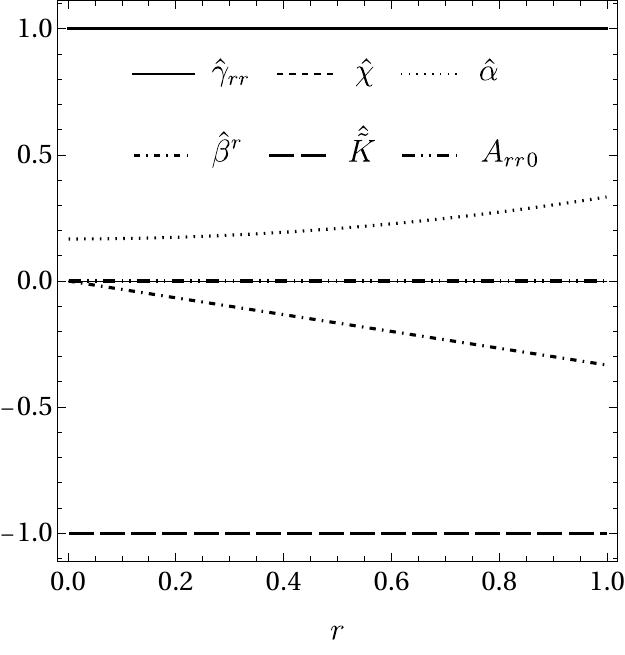}
\caption{Reference metric ``CMC'': rescaled height function $h$ \eref{cmch} and compactification factor $\aconf$ \eref{cmcaconf} on the left, physical propagation speeds $c_\pm$ \eref{speeds} on the conformally compactified hyperboloidal slice in the center, and relevant reference metric quantities \eref{cmchatvals} and initial value for $A_{rr}$ on the right. The requirement $\hat{\gamma}_{rr}=1$ has been used here. Compare to the depiction of other reference metrics in \aref{pics}.}
\label{cmcfigs}
\end{figure}

A visual representation of the different parts of the construction is presented in \fref{cmcfigs}. On the left, the height function is shown rescaled by the compactification factor because $h(r=\rscri)=\infty$. However, $\aconf\,h|_{r=\rscri}=\rscri=1$, which accounts for a more convenient visualization. The central plot displays the physical propagation speeds of the system 
\begin{equation}\label{speeds}
c_\pm=-\beta^r\pm\alpha\sqrt{\frac{\chi}{\gamma_{rr}}}
\end{equation}
for the chosen reference metric on the conformally compactified hyperboloidal slice. 
Finally, the profiles of the components of the reference metric, the value of the corresponding trace of the physical extrinsic curvature, and the initial value for $A_{rr}$ are included in the right plot. These depictions are to be compared to those of other reference metrics included in \aref{pics}.

\subsection{Height function with unity outgoing propagation speed -- labelled as ``$c_+=1$''}\label{cplusss}

The height function commonly used in the dual frame approach \cite{Hilditch:2015qea,Hilditch:2016xzh} to free hyperboloidal evolutions \cite{Gasperin:2019rjg,Gautam:2021ilg,Peterson:2023bha} is chosen \cite{Zenginoglu:2010cq} such that the outgoing propagation speed of the system is unity, $c_+=1$. This translates to the expression $\partial_{\tilde r}h = 1-\left(\frac{d\tilde r}{dr}\right)^{-1} = 1-\frac{dr}{d\tilde r}$, which after integrating in $r$ and substituting \eref{trafor} gives 
\begin{equation}\label{cplush}
h(r) = \frac{r}{\aconf}-r . 
\end{equation}
For this height function, also called ``characteristic-preserving'' in \cite{Zenginoglu:2024bzs}, $c_+=1$ will hold (see central plot in \fref{cplusfigs}) for any compactification $\aconf$. However, this means that the choices of slicing and compactification are interrelated, and changing $\aconf$ will change the slice. Moreover, within the present approach it is not feasible to impose an explicitly conformally flat spatial metric satisfying \eref{aconfode}, because the physically-meaningful positive solution to the latter yields $\aconf=\log(r)/(\log(r)-2)$. This expression, while being finite at the origin, has a diverging derivative there, and thus cannot be used for numerical evolutions including $r=0$.  
Instead, the compactification factor is chosen as in \cite{Hilditch:2016xzh,Gasperin:2019rjg,Gautam:2021ilg,Peterson:2023bha}, which coincides with \eref{cmcaconf} for $\K=-6/\rscri$ and is unity at the origin, 
\begin{equation}\label{cplusaconf}
\aconf=1-\frac{r^2}{\rscri^2} . 
\end{equation}
The same quantities as displayed in \fref{cmcfigs} are shown for the ``$c_+=1$'' case in \fref{cplusfigs}. One further relevant aspect to note here is that the height function \eref{cplush} has odd parity at the origin (and its derivative along the compactified radial coordinate is even there). The consequence of this is that the radial component of the shift $\beta^r$ constructed from \eref{refmetcomp} is even at the origin. This is not the expected behaviour of the shift, and attempting evolutions with the present reference metric required modifying the parity conditions for $\beta^r$ at the origin in the numerical implementation. The conformal compactification method to the Einstein equations requires a height function that is even at the origin to provide the correct parity for the evolution variables, so that this ``$c_+=1$'' choice together with \eref{cplusaconf} is not optimal. 

\subsection{Boost ($\partial_{\tilde r}h$) defined in terms of the compactified radial coordinate -- labeled ``cos'' and ``sin''}\label{cossinss}

The two cases considered here (labeled as ``cos'' \eref{coshp} and ``sin'' \eref{sinhp}) are similar to the previous one in that the slicing and the compactification are entwined. Also, the condition $\hat\gamma_{rr}=1$ cannot be satisfied by these choices, as will be explained later on. 

The ``cos'' reference metric is based on 
\begin{equation}\label{coshp}
\partial_{\tilde r} h= \frac{1-\cos(\pi r/\rscri)}{2}, 
\end{equation}
selected following the requirements that $\partial_{\tilde r}h|_{r=0}=0$, $\partial_{\tilde r}h|_{r=\rscri}=1$, and $0<\partial_{\tilde r}h<1$ always in-between. 
As $\partial_{\tilde r}h$ is  defined  in  terms  of $r$, a choice of $\aconf$ must be made to calculate $h$. One may integrate either
\begin{equation}\label{hdersrel}
\partial_{\tilde r}h(r=\tilde r\,\aconf(\tilde r))\qquad \textrm{or}\qquad \partial_{r}h(r) = \frac{d\tilde r}{dr}\left[\partial_{\tilde r}h\right](r) = \frac{(\aconf-r\aconf')}{\aconf^2} \left[\partial_{\tilde r}h\right](r) . 
\end{equation}
The integrated height function $h$ is not needed for the simulations' reference metric or initial data, but it is $h$ that determines the slices, so that the choice of compactification factor will necessarily influence them. Also, the integrated $h$ is needed to create the conformal diagrams in \fref{penfig}. 

Satisfying $\hat\gamma_{rr}=1$ with \eref{coshp} is not possible, because the expression chosen for $\partial_{\tilde r}h$ is given exclusively in terms of the compactified $r$ (and not $\tilde r$). This means that $\aconf$ does not appear in the part in parenthesis in \eref{aconfode}, and $\aconf'\propto\aconf/r$ does not contain a meaningful solution for $\aconf$ that satisfies $\aconf(\rscri)=0$ and $\aconf(r<\rscri)>0$. The same is valid for option ``sin'' \eref{sinhp} considered later in this subsection. 

The CMC choice \eref{cmcaconf} is made here for the compactification $\aconf$ and conformal $\Omega$ factors. Integration of \eref{coshp} using the second relation in \eref{hdersrel} yields
\begin{equation}\label{cosh}
h(r) = \frac{-3}{2\K}\left(\frac{2\, r\, \rscri \left(1-\cos \left(\frac{\pi  r}{\rscri}\right)\right)}{\rscri^2-r^2}+\pi  \left(\text{Si}\left[\pi\left(1-\frac{r}{\rscri}\right)\right]-\text{Si}\left[\pi\left(1+\frac{r}{\rscri}\right)\right]\right)\right),
\end{equation}
where $\text{Si}[z]=\int_0^z\sin(x)/x\,dx$ is the sine integral function. This height function is odd at the origin, as could be expected from its radial derivative \eref{coshp} being even. Thus, it will provide the wrong parity at the origin for some of the evolution quantities in this approach. The rescaled profile of \eref{cosh}, the corresponding reference metric and propagation speeds are shown in \fref{cosfigs}. 

For the ``sin'' reference metric, an odd radial derivative of the height function is chosen as
\begin{equation}\label{sinhp}
\partial_{\tilde r} h= \sin\left(\frac{\pi\,r}{2\,\rscri}\right) . 
\end{equation}
Following the same integration procedure as above using the CMC \eref{cmcaconf} compactification factor, the corresponding height function is given by
\begin{align}\label{sinh}
h(r)=&\frac{-3}{2 \K} \left(\frac{4\, r\, \rscri \sin \left(\frac{\pi  r}{2 \rscri}\right)}{\rscri^2-r^2}+\pi  \left(\text{Si}\left[\frac{\pi}{2}\left(1-\frac{r}{\rscri}\right)\right]+\text{Si}\left[\frac{\pi}{2}\left(1+\frac{r}{\rscri}\right)\right]\right)\right) \nonumber \\&+ \frac{3 \pi\, \text{Si}\left(\frac{\pi }{2}\right)}{\K} , 
\end{align}
where the last term has been added to satisfy $h(r)=0$. This height function is even at the origin, which provides the correct parity for the reference metric and evolution variables. These are depicted, together with other relevant quantities, in \fref{sinfigs}. 

\subsection{Height function similar in shape to CMC -- labeled as ``sim''}\label{simss}

The ``sim'' reference metric uses the height function (16) in \cite{Zenginoglu:2024bzs}, subtracting $C$ to make it zero at the origin: 
\begin{equation}\label{simh}
h(\tilde r) = \tilde r+\frac{C}{1+\tilde r}-C \quad\longleftrightarrow\quad h(r) = \frac{r}{\aconf}+\frac{C}{1+r/\aconf}-C . 
\end{equation}
Its shape looks qualitatively similar to the CMC one -- see the $\partial_{\tilde r}h$ profiles in \fref{boostsum} for comparison. However, the ``sim'' height function has mixed parity at the origin. This comes out most clearly in the profile of $\hat{\tilde K}$ on the right plot in \fref{simfigs}, where it is neither even nor odd there. This property makes a reference metric choice based on \eref{simh} not ideal for the numerical implementation. 

An advantage of this height function is that it allows to impose $\hat\gamma_{rr}=1$ after substituting the derivative of \eref{simh} into \eref{aconfode}, and solving for $\aconf$ numerically. The result is shown in the left plot in \fref{simfigs} together with the rescaled height function. The speeds and reference variables can be seen on the central and right graphs. 

\subsection{Modification of CMC height function -- labeled as ``mod'', with variation labeled as ``modvar''}\label{modss}

This modification (``mod'') takes the form of a zero-at-origin, even-parity version of the second term in \eref{simh} multiplying the squared-radius term in the CMC height function \eref{cmch}: 
\begin{subequations}
\begin{align} \label{modh}
h(\tilde r) = \sqrt{\left(\frac{3}{\K}\right)^2+\tilde r^2\frac{\tilde r^2}{C^2+\tilde r^2}}+\frac{3}{\K}\quad\longleftrightarrow\quad  h(r) = \sqrt{\left(\frac{3}{\K}\right)^2+\left(\frac{r}{\aconf}\right)^2\frac{(r/\aconf)^2}{C^2+(r/\aconf)^2}}+\frac{3}{\K} . 
\end{align} 
\end{subequations}
The effect of the modification is to make the height function less steep at the origin, while maintaining the desired parity there. For $C=0$, the CMC expression is recovered. One limitation of this ansatz is that $C$ must be smaller than 3, as above that value of the parameter $\partial_{\tilde r}h>1$ near \scrip. This means that the slices become timelike in a region of spacetime, which is not the required behaviour. 
This height function allows to set $\hat\gamma_{rr}=1$ for \eref{modh}, and obtain from \eref{aconfode} a numerical solution for $\aconf$, shown on the left in \fref{modfigs} for $C=2.8$. As $C$ is chosen to be closer and closer to 3, the decay of the numerically-found $\aconf$ becomes steeper and steeper there, meaning that its first derivative diverges as $C\to 3$. 

Alternatively, instead of the numerically solved one, the CMC compactification factor \eref{cmcaconf} can be set for $\aconf$ and $\Omega$; in this case $\hat\gamma_{rr}\neq 1$. The profiles of rescaled $h$, propagation speeds and reference quantities are shown for $C=2$ in \fref{modvarfigs}. This variation, in which the slice remains the same but the compactification is different, is called ``modvar''. 

\subsection{CMC-modeled hyperboloidal layer -- labeled as ``layCMC''}\label{layalmss}

The starting point is the boost function ($\partial_{\tilde r}h$) (33) in \cite{Zenginoglu:2010cq}, which in the present notation gives the height function
\begin{subequations}\label{layalmh}
\begin{align}
h(\tilde r)=\left[\sqrt{\left(\frac{3}{\K}\right)^2+\left(\tilde r-\rmatch\right)^2}+\frac{3}{\K}\right]\Theta(\tilde r-\rmatch) \quad \longleftrightarrow \\
h(r)=\left[\sqrt{\left(\frac{3}{\K}\right)^2+\left(\frac{r}{\aconf}-\rmatch\right)^2}+\frac{3}{\K}\right]\Theta(r/\aconf-\rmatch) ,
\end{align} 
\end{subequations}
with $\rmatch$ the location in the uncompactified radial coordinate where the matching takes place and $\Theta$ the Heaviside step function. An alternative to using the compactification function (32) in \cite{Zenginoglu:2010cq} is to substitute the present \eref{layalmh} into \eref{aconfode} and solve for $\aconf$, which provides the closed-form solution
\begin{subequations}
\begin{align}
&\aconf(r)=-\frac{2 \K r \cosh ^2\left(\frac{1}{6} \sqrt{\K^2 \rmatch^2+9} \left[\log (\rscri)-\log (r)\right]+\tanh ^{-1}\left(\frac{\K \rmatch}{\sqrt{\K^2 \rmatch^2+9}}\right)\right)}{\left(\K^2 \rmatch^2+9\right)^{3/2}} \times \nonumber\\
&\left[\left(\K^2 \rmatch^2+9\right) \tanh \left(\frac{1}{6} \sqrt{\K^2 \rmatch^2+9} \left[\log (\rscri)-\log (r)\right]+\tanh ^{-1}\left(\frac{\K \rmatch}{\sqrt{\K^2 \rmatch^2+9}}\right)\right) \right. \nonumber\\ &\left. -\K \rmatch\sqrt{\K^2 \rmatch^2+9}\right] \qquad \textrm{for}\quad r\ge \rcompmatch,\label{layaconf} \\ 
&\textrm{with} \quad \rcompmatch = \, \rscri \exp \left(-\frac{6 \cosh ^{-1}\left(\frac{3-\K \rmatch}{\sqrt{6} \sqrt{-\K \rmatch}}\right)}{\sqrt{\K^2 \rmatch^2+9}}\right) , \  \textrm{from} \  \frac{\rcompmatch}{\aconf(\rcompmatch)}=\rmatch . 
\end{align}
\end{subequations}
This satisfies $\hat\gamma_{rr}=1$ and reduces to \eref{cmcaconf} for $\rmatch=0$. In the Cauchy part where $r<\rcompmatch$, the value $\aconf(\rcompmatch)$ is to be set. 
Putting \eref{layaconf} together with \eref{layalmh}, the resulting compactified hyperboloidal layer and corresponding reference metric are continuous at the matching point. They are however not smooth there, so that the components of the extrinsic curvature present a jump, making this setup unusable numerically. Using the compactification (32) in \cite{Zenginoglu:2010cq} (less smooth than \eref{layaconf}) provides similar non-smooth results. 

The ``layCMC'' reference metric setup builds from the radial derivative of \eref{layalmh}, applying a smooth matching function to ensure all quantities are continuous at the matching point. The matching function is taken to be
\begin{equation}
f_{match}=\left[1-\left(\frac{\tilde r_+^2-\tilde r^2}{\tilde r_+^2-\tilde r_-^2}\right)^4\right]^4 , 
\end{equation}
and set into the now smoother expression 
\begin{equation}\label{laysmooth}
\partial_{\tilde r}h(\tilde r)=\left\{\begin{array}{llc}
0&\textrm{if}&\tilde r\le\tilde r_- \\ 
0\times(1-f_{match}) + f_{match}\times\frac{\left(\tilde r-\rmatch\right)}{\sqrt{\left(\frac{3}{\K}\right)^2+\left(\tilde r-\rmatch\right)^2}} &\textrm{if}&\tilde r_-<\tilde r<\tilde r_+\\ 
\frac{\left(\tilde r-\rmatch\right)}{\sqrt{\left(\frac{3}{\K}\right)^2+\left(\tilde r-\rmatch\right)^2}}&\textrm{if}&\tilde r_+\le\tilde r\end{array}\right. . 
\end{equation}
The chosen matching parameters are $\rmatch=5$, $\tilde r_-=\rmatch$ and $\tilde r_+=\rmatch+2.5$. Substituting also $\K=-1$ and $\rscri=1$, \eref{aconfode} can be numerically solved for a correction to \eref{layaconf} that makes $\hat\gamma_{rr}=1$ hold everywhere. 
\Fref{layalmfigs} displays the rescaled height function (integrated from \eref{laysmooth}) and the obtained compactification, as well as the propagation speeds and reference metric components corresponding to the hyperboloidal layer setup found. 
As can be observed in the right-most plot, all quantities are continuous and differentiable at the matching point, located around $r=0.4$. No boundary conditions are required at the matching point between Cauchy and hyperboloidal parts, as the slice is continuous and spacelike everywhere. Numerically, it is treated as a single slice with gridfunction values given piecewise by \eref{laysmooth} and its corresponding compactification. The same is valid for the hyperboloidal layer setups that follow. 

\subsection{Layer modeled with ``mod'' height function -- labeled as ``laymod'', with variation labeled as ``layvar''}\label{laymodss}

Here the ``mod'' height function \eref{modh} is put together with the hyperboloidal layer construction. The required adaptation is an off-centering of the origin of the original function, yielding
\begin{subequations}\label{laymodh}
\begin{align} 
h(\tilde r) = \left[\sqrt{\left(\frac{3}{\K}\right)^2+(\tilde r-\rmatch)^2\frac{(\tilde r-\rmatch)^2}{C^2+(\tilde r-\rmatch)^2}}+\frac{3}{\K}\right]\Theta(\tilde r-\rmatch)\quad\longleftrightarrow\\ 
h(r) = \left[\sqrt{\left(\frac{3}{\K}\right)^2+\left(\frac{r}{\aconf}-\rmatch\right)^2\frac{(r/\aconf-\rmatch)^2}{C^2+(r/\aconf-\rmatch)^2}}+\frac{3}{\K}\right]\Theta(r/\aconf-\rmatch) . 
\end{align} 
\end{subequations}
The behaviour of this ansatz near the matching point is smooth enough to provide suitable initial data for all quantities, so no further smoothing is required. 

The ``laymod'' reference metric uses the height function above, and solves \eref{aconfode} for a correction to the compactification factor \eref{layaconf} with $\rmatch=4$, $C=2$, $\K=-1$ and $\rscri=1$. (The same process as in the previous subsection). This makes the spatial reference metric explicitly conformally flat. The results for all quantities are included in \fref{laymodfigs}. 

For the ``layvar'' reference metric, $\hat\gamma_{rr}\neq 1$ and the compactification factor is chosen as \eref{layaconf} with the same parameter choices as for ``laymod''. In principle, one would want the Cauchy part to be completely unaffected by the hyperboloidal layer, and thus having $\Omega=\aconf=1$ in the interior would be ideal. Given the present choice of $\rscri$ and $\rmatch$ that is not possible -- $\rmatch<\rscri$ must hold for it to be feasible. To better compare with the other reference metrics, the values of $\rscri$ and $\rmatch$ are maintained, and further experimentation on this is left for future work. The profiles of the reference metric and related quantities are shown in \fref{layvarfigs}. 

\subsection{Visualization}\label{visu}

\begin{figure}[h]
\centering
\includegraphics[width=0.99\textwidth]{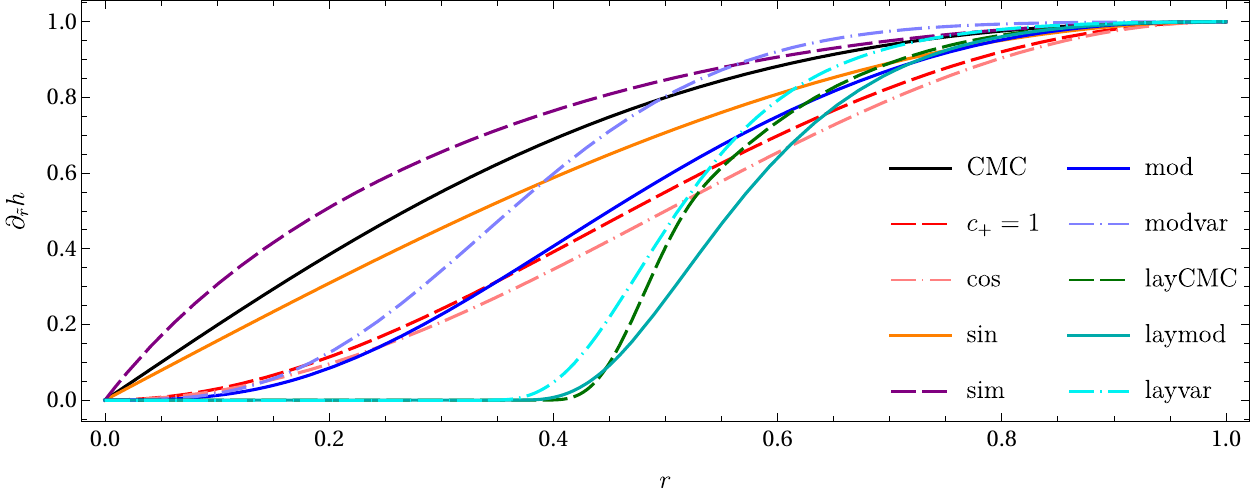}
\caption{Summary of all $[\partial_{\tilde r}h](r)$ considered, with the same legend as used in \fref{monitorfigs}.}
\label{boostsum}
\end{figure}

\begin{figure}[h]
\centering
\includegraphics[width=0.24\textwidth]{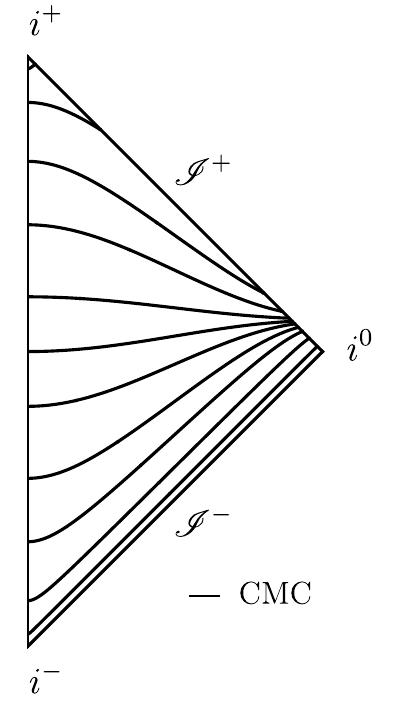}
\includegraphics[width=0.24\textwidth]{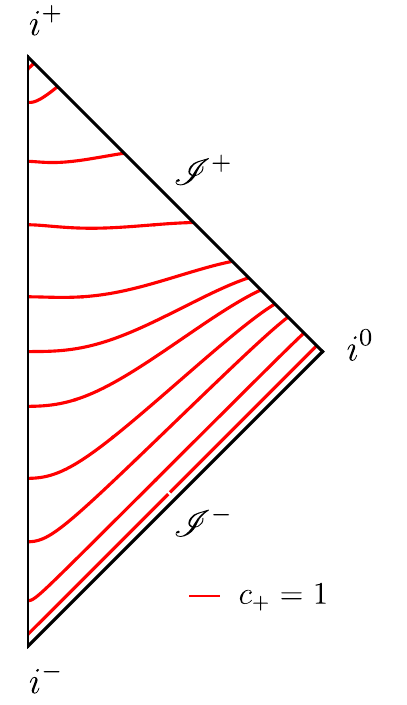}
\includegraphics[width=0.24\textwidth]{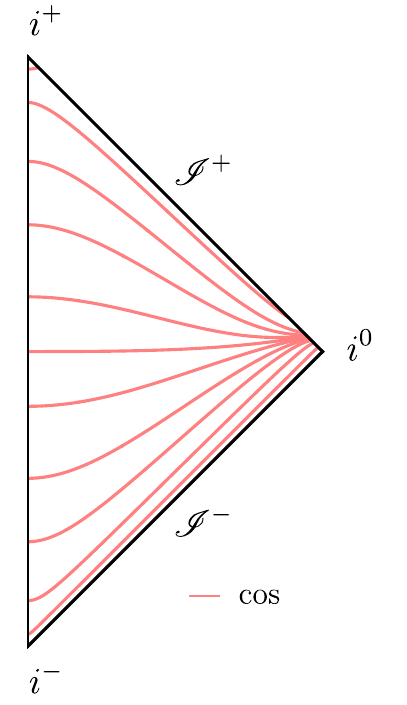}
\includegraphics[width=0.24\textwidth]{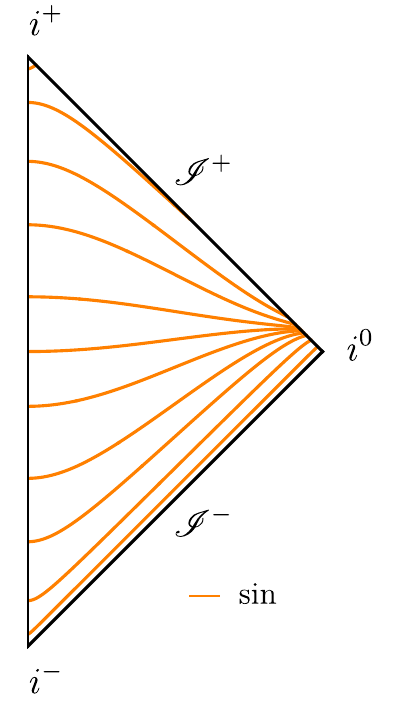}
\includegraphics[width=0.24\textwidth]{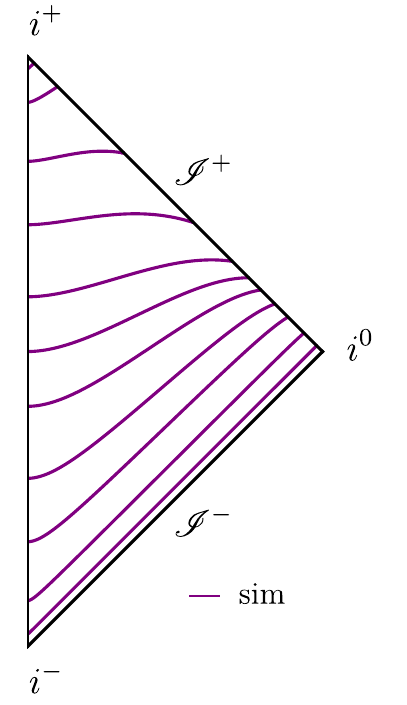}
\includegraphics[width=0.24\textwidth]{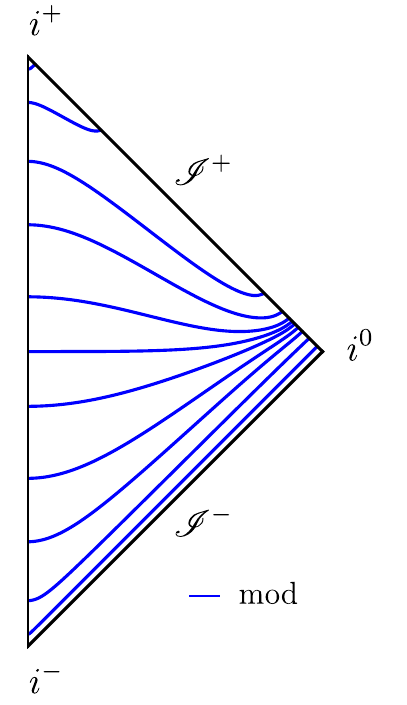}
\includegraphics[width=0.24\textwidth]{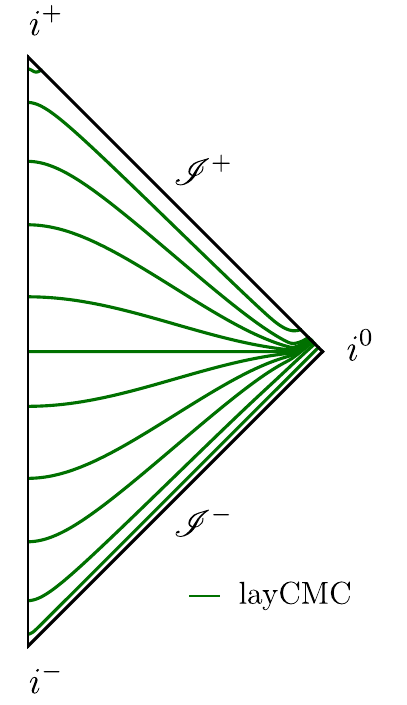}
\includegraphics[width=0.24\textwidth]{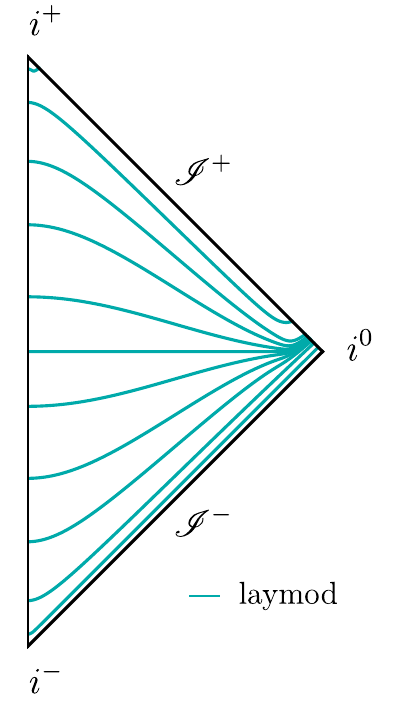}
\caption{Penrose diagrams of the distinct slices considered. The ``mod'' and ``modvar'' slices are the same, as well as the ``laymod'' and ``layvar'' ones, because what changes between each pair is just the conformal compactification.} 
\label{penfig}
\end{figure}

\Fref{boostsum} and \fref{penfig} are intended as a summary of the hyperboloidal slices of Minkowski spacetime described in the previous subsections. In \fref{boostsum}, $\partial_{\tilde r}h$ as a function of the compactified radial coordinate $r$ is shown for all height functions considered. All profiles take monotonically increasing values between 0 and 1, as expected. 
In \fref{penfig}, most considered hyperboloidal slices are depicted in conformal Carter-Penrose diagrams. Construction of the conformal diagrams is performed in the usual way, see e.g. \cite{\pappen}. The slicing is determined by the height function $h$. The choice of compactification $\aconf$ or conformal $\Omega$ factor for the reference metric data does not play a role in these diagrams\footnote{The choice of compactification \textit{along the null directions} for the construction of the conformal diagrams, chosen to be $\arctan$ for all cases here as in \cite{\pappen}, does have an impact on the curves, but is irrelevant for the numerical experiments of interest in this work.}. The ``cos'' case is not shown because it is visually indistinguishable from the ``sin'' one. The ``mod'' and ``modvar'' height functions are exactly the same, with only the compactification and conformal rescaling changing between both setups. The same holds for ``laymod'' and ``layvar'', which look very similar to ``layCMC''. Note in the latter's diagram how the slices look like $\tilde t=constant$ Cauchy slices in the interior, and only very close to \scrip do they tilt upwards. 

\section{Results}\label{resu}

The 1+1D code solving the conformally rescaled Einstein equations in spherical symmetry for this work uses the method of lines with 4th-order finite differences for the spatial derivatives and a Runge-Kutta of 4th order as time integrator. The advection terms are upwinded in the usual way to increase numerical stability. The added Kreiss-Oliger dissipation \cite{kreiss1973methods} uses a strength of $\epsilon=0.1$ for these simulations. The values of the parameters used in the evolutions are $\kappa_1=1.5$ and $\kappa_2=0$ for the constraint damping terms (with the $\kappa_1$ term present in both BSSN and Z4c), $\cnK=1.5$, $\nconst=1$, $\eta=0$, $\lambda=1$, $\K=-1$ and $\rscri=1$. The parameter $C$ takes the value of 2.8 for the ``mod'' simulations, and of 2 in the other cases when present. 

\begin{figure}[h]
\centering
\includegraphics[width=0.49\linewidth]{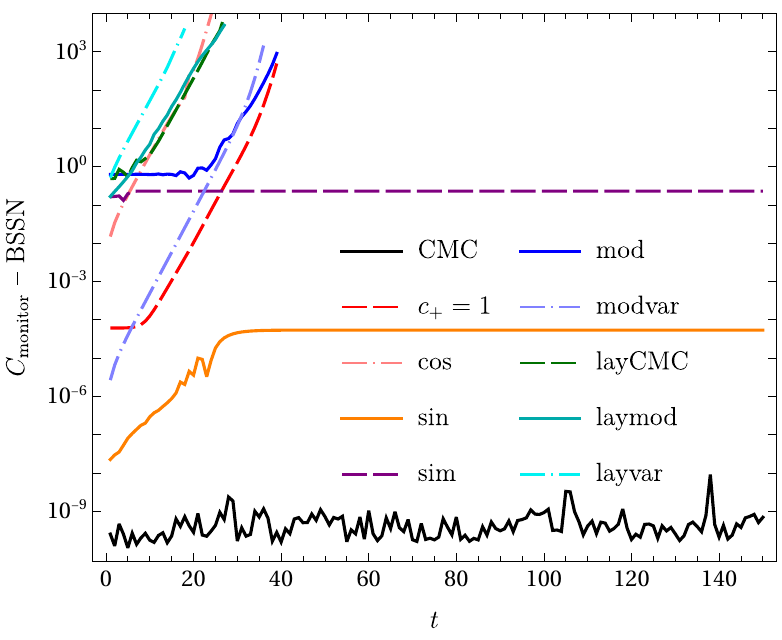}
\includegraphics[width=0.49\linewidth]{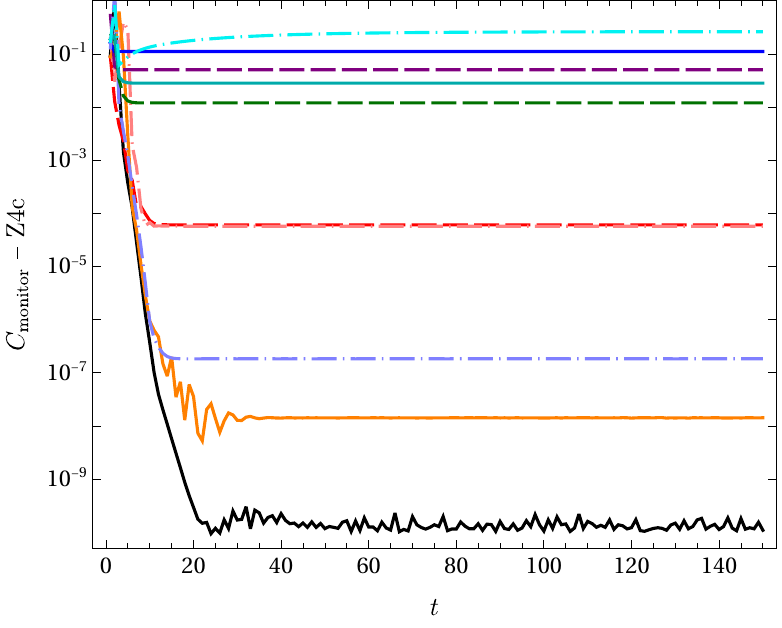}
\caption{Long-term behaviour of the simulations as illustrated by the monitoring function \eref{fullmonitor}. The legend is the same in both plots and in \fref{boostsum}. The left graph shows that only the CMC, sin and sim reference metrics provide a long-term stable evolution with the BSSN equations. However, using Z4c gives stable simulations for all choices of reference metric considered, as depicted on the right for initial data with superimposed noise of $\sim 10^{-4}$ amplitude.}
\label{monitorfigs}
\end{figure}

The objective of this study is to determine which reference metrics provide a long-term stable evolution. What is meant here with long-term stability is that at late times in the simulation the system has found a numerical solution and does not deviate from it. 
One way to test this type of long-term behaviour is via a robust stability test \cite{Babiuc_2008}. It basically consists of adding random noise to initial data that coincides with the final stationary solution of the system. The results will be considered stable if the solution is below a finite upper bound in some suitable norm. Drawing inspiration from the robust stability test, random noise with an amplitude of $10^{-4}$ is added to the initial data for the simulations using the Z4c equations. This large noise amplitude has been chosen to put a stringent test on the data, and also to make it more noticeable in the plot (right graph in \fref{monitorfigs}).

Evolutions are run with each of the 10 reference metrics considered (CMC, $c_+=1$, cos, sin, sim, mod, modvar, layCMC, laymod, layvar). The reference metric components are set in the gauge source functions in \eref{slicing} and \eref{shift}, and coinciding initial data are set for the system. The expectation is therefore that no dynamics should take place during the simulation if the system is long-term stable. The evolutions are run until $t=150$ with the conformally rescaled BSSN and Z4c equations, using 400 spatial gridpoints for $r\in(0,1)$ and a timestep of $dt=0.0005$. The results are summarised in \fref{monitorfigs}. 

A convenient way to monitor the behaviour of the simulations is by looking at the behaviour of their constraints over time. Expression (73) in \cite{Cao:2011fu} is useful for that, and here has been adapted to
\begin{subequations}
\begin{align}
C_{\textrm{monitor}}=& \sqrt{\frac{1}{4\pi}\int\left(H^2+\bar\gamma^{ij}M_iM_j+\gamma^{ij}Z_iZ_j+\kappa_{Z4}\tilde\Theta^2\right)\sqrt{\gamma}\,dV} \\=& \sqrt{\int_0^{\rscri}\left(H^2+\frac{\chi\,M_r^2}{\gamma_{rr}}+\frac{Z_r^2}{\gamma_{rr}}+\kappa_{Z4}\tilde\Theta^2\right)r^2dr} , 
\end{align}
\end{subequations}
where $\kappa_{Z4}=0$ for BSSN and 1 for Z4c. The continuum expression above 
takes the following form in terms of the discrete gridvectors evaluated on the $N_{\textrm{points}}$ spatial points
\begin{equation}\label{fullmonitor}
C_{\textrm{monitor}}= \sqrt{\frac{\sum_{i=1}^{N_{\textrm{points}}}\left[H_i{}^2+M_{r}{}_i^2\chi_i/\gamma_{rr}{}_i+Z_{r}{}_i^2/\gamma_{rr}{}_i+\kappa_{Z4}\tilde\Theta_i{}^2\right]r_i^2}{N_\textrm{points}}} , 
\end{equation}
and provides a useful way to compare the long-term behaviour of the simulations, shown in \fref{monitorfigs}. The left plot shows the results for BSSN (where no noise was added to the initial data), while the right one presents the Z4c evolutions with noise-perturbed initial data. Curves that become horizontal towards later times indicate that an upper bound exists for the solution and they are long-term stable. This is the case for all simulations run with Z4c and the ``CMC'', ``sin'' and ``sim'' cases run with BSSN. All other configurations evolved with BSSN were unstable: a continuum instability would develop slowly, cause the variables to drift away from their initial data, and finally make the simulation crash. This is illustrated by the lines at the top-left corner of the left plot in \fref{monitorfigs}. Given the unstable behaviour of most simulations run with BSSN, no noise was added to the initial data. For ``sim'', one the surviving simulations, there were relevant constraint violations at late times for the stationary solution found. This may have been related to the mixed parity conditions at the origin not dealt with fully consistently. In order to test all setups on a common ground, the parameters were not tuned to each of them -- doing so may have rendered some of the unstable simulations better behaved. Further studies on this are left for future work. 

The Z4c evolutions (right plot in \fref{monitorfigs}) gave very different results: all reference metric choices successfully arrived at a long-term stable stationary configuration. This is most likely due to the constraint propagation properties of the Z4 system, which drive numerical error away instead of letting it accumulate in the same coordinate location. To test robustness of the simulations, $10^{-4}$-amplitude noise was added to the initial data, and it was dissipated away successfully. For some cases, the constraint violations of the long-term stationary solution found are rather large. This is especially the case for ``layvar'', where the Hamiltonian constraint would take a non-vanishing value at the origin and extending to the Cauchy part of the slice at late times. This seems to improve as resolution is increased, but a more thorough investigation is left for the future. The first attempt with ``modvar'' (height function \eref{modh} and CMC compactification \eref{cmcaconf}) used $C=2.8$, but it crashed at \scrip very fast. Setting $C=2$ (as in \fref{modvarfigs} and \fref{monitorfigs}) worked -- this is probably related to the difference in behaviour near \scrip between the height function and the chosen compactification factor, pointing to a subtle equilibrium between quantities near future null infinity. 

The control case ``CMC'' behaved the same in simulations with BSSN and Z4c: the black curves in \fref{monitorfigs} just show noise at the level of machine precision in the constraints. The final constraint violations for ``sim'' are much smaller in the Z4c case than in the BSSN one. This is also the case for the ``sin'' reference metric, although the difference is not as pronounced. Only the Z4c system was able to deal with the matching point in the hyperboloidal layer configurations; the runs with BSSN started showing problems exactly at the joints between Cauchy and hyperboloidal parts, which again seems to indicate that successfully propagating and damping the errors plays a very important role. Finally, the convenient condition $\hat\gamma_{rr}=1$ is not required for stability: the ``sin'' case, whose spatial reference metric is not explicitly conformally flat, was long-term stable with both BSSN and Z4c formalisms. 

\begin{figure}[h]
\centering
\includegraphics[width=0.32\textwidth]{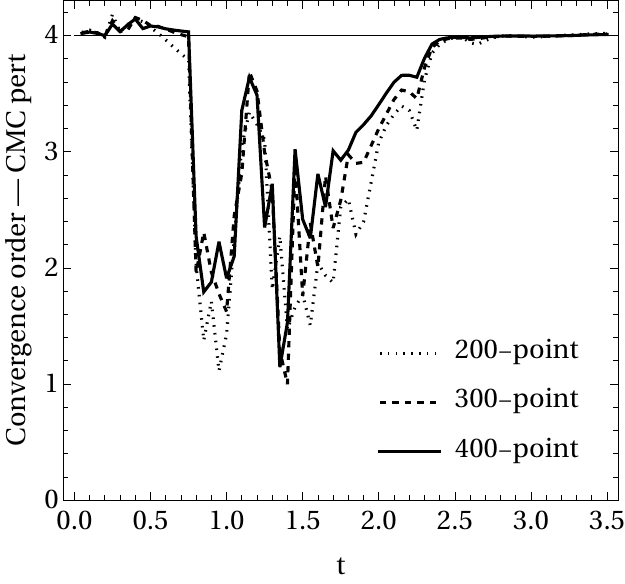}
\includegraphics[width=0.32\textwidth]{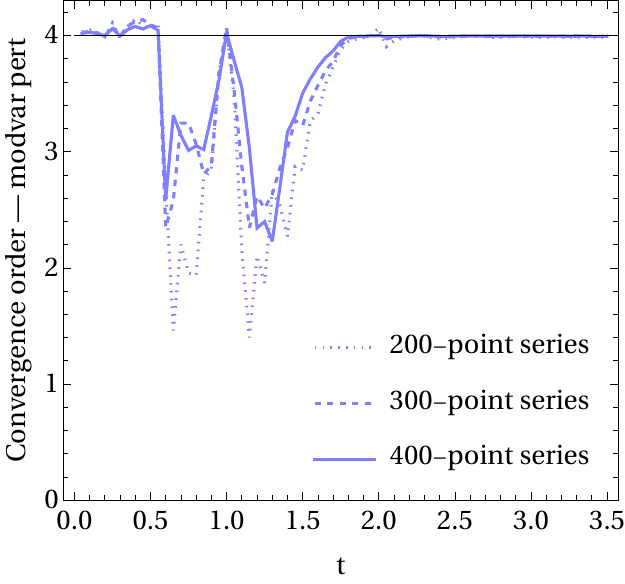}
\includegraphics[width=0.325\textwidth]{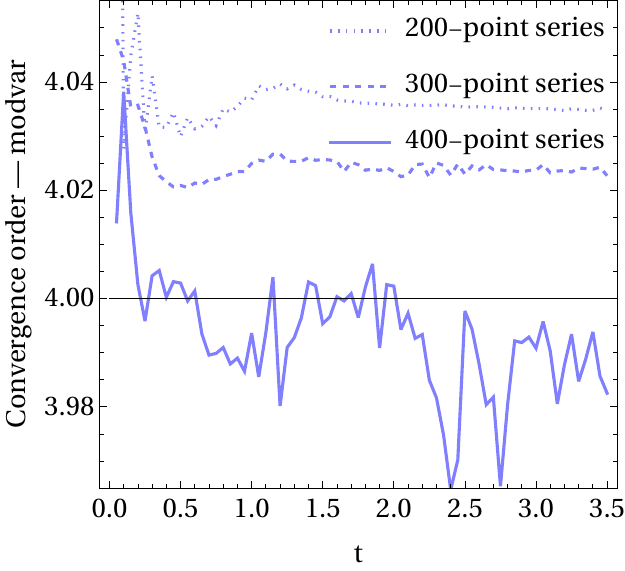}\\
\includegraphics[width=0.64\textwidth]{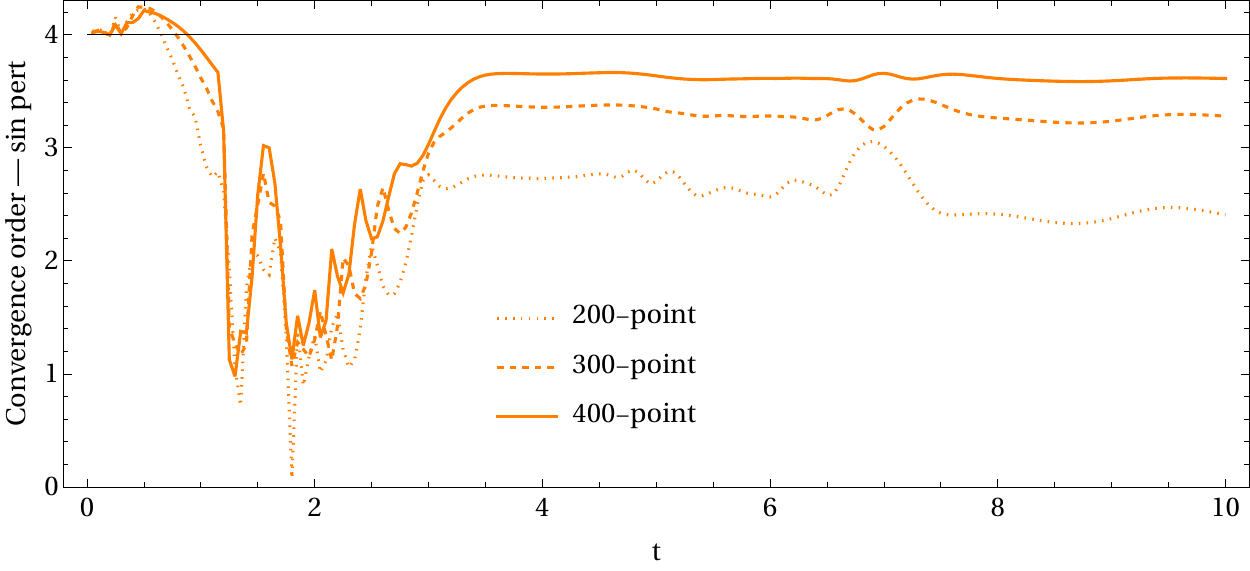}
\includegraphics[width=0.34\textwidth]{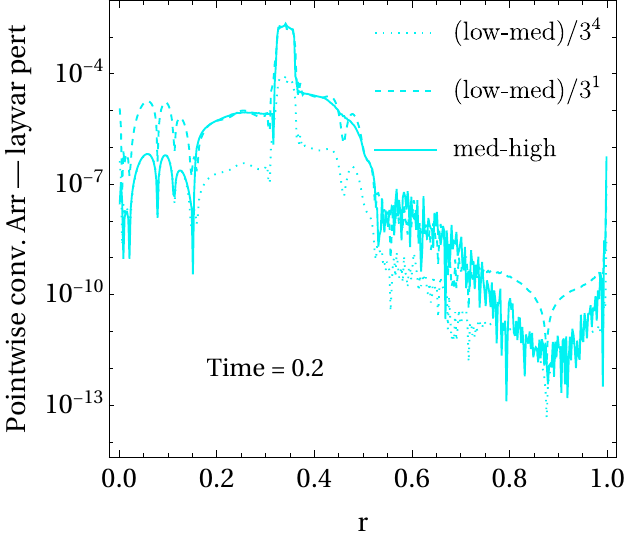}
\caption{Plots showing the convergence of some of the setups considered. The convergence order over time as shown on the top row and bottom-left plot is calculated using the $L^2$-type norm as $\log_f\left(\frac{\sqrt{\sum_{i=1}^N(X_{low, i}-X_{med, i})^2}}{\sqrt{\sum_{i=1}^N(X_{med, i}-X_{high, i})^2}}\right)$ with $i\in[1,N]$. The increase in resolution is $f=3$, the number of points $N=200,300,400$ for each of the series, and $X_{low/med/high, i}$ denotes the evolution variables for the low, medium and high resolutions at point $i$. All the evolution variables were included in the sums, so that $X$ was successively $\chi,\gamma_{rr},A_{rr},\Delta\tilde K,\Lambda^r,\alpha,\beta^r$ and $\tilde\Theta$. The ``CMC pert'' (top-left), ``modvar pert'' (top-center) and ``sin pert'' (bottom-left) cases simulate a small gauge perturbation whose behaviour at the origin and \scrip causes the deviations from the expected 4th order convergence, which is much better at later times when the perturbation has left the domain. The top-right ``modvar'' simulations include no perturbation and show very good 4th order convergence. The bottom-right plot shows a pointwise convergence plot for the ``layvar pert'' case at time 0.2 for the evolution variable $A_{rr}$, with a logarithmic scale on the $y$ axis for ease of visualization. At this time the expected convergence order is still maintained near the origin and near \scrip, as the solid line ($A_{rr}{}_{med, i}-A_{rr}{}_{high, i}$) coincides there with the dotted one ($A_{rr}{}_{low, i}-A_{rr}{}_{med, i}$ rescaled by $3^4$ assuming 4th order convergence). However, around $r=0.35$ the solid line coincides with the dashed one, for whose rescaling 1st order convergence was assumed. See main text for further details.}
\label{convfig}
\end{figure}

The usual way in numerical relativity to show that the results obtained are reliable is by performing a convergence test. \Fref{convfig} shows plots for four different setups including three sets of convergence runs each. The latter are series with spatial resolutions of 200-600-1800, 300-900-2700 and 400-1200-3600 points, with timesteps rescaled accordingly. The fifth plot (bottom-right) only shows data from a higher resolution set. The resolution increase between runs in each series is 3, because for a staggered grid (or cell-centered grid, meaning $r=0$ and $r=\rscri=1$ fall between gridpoints) as used here, the points for different resolutions align and interpolation is not needed for checking convergence (see figure 6.3 in \cite{\thesis} for the exact setup). 

The top-left, top-center and bottom-left norm convergence plots in \fref{convfig} correspond respectively to simulations of the ``CMC'', ``modvar'' and ``sin'' cases that include a small Gaussian-like perturbation superimposed on the initial data for the lapse. The simulations on the top-right for ``modvar'' do not include the perturbation. The ideal outcome for the norm convergence tests is a horizontal line at 4, for the 4th order convergence of the finite differences and time integrator used. A expected, the largest deviations from 4 appear in the dotted lines (lowest resolution set further away from the exact solution). The case without perturbation on the right shows indeed very good convergence, which gets better for the 3rd convergence series (solid line) with higher resolution. The same setup (without initial perturbation) was run with ``CMC'' and ``sin'', but the convergence did not look good: this is not because the results are not reliable, but because the differences between the variables were too small to show convergence properly. The top-left, top-center and bottom-left plots with initial gauge perturbation show clearly that the convergence order is recovered better as the resolution of the series increases, but there is still a region where all lines deviate considerably from 4, and in a similar way for all three considered reference metrics. This loss of convergence coincides with the arrival of the gauge perturbations at \scrip, and is due to the chosen gauge conditions. An equivalent effect (for a black hole spacetime) for the ``CMC'' case can be observed in Fig. 14 in \cite{\papbh}. While this behaviour is far from ideal, the gauge condition studies required for its improvement fall beyond the scope of this work.  

The ``sin pert'' simulations whose norm convergence is shown in the bottom-left plot in \fref{convfig} were run for longer to observe the long-term behaviour of convergence, as the curves do not immediately go back to 4 after the gauge perturbation has left the domain, unlike the ``CMC'' and ``modvar'' cases shown. Pointwise self-convergence plots (equivalent to that shown on the bottom-right in \fref{convfig}) of the evolution variables at $t\sim 4$ show that the rescaled differences at a given gridpoint and time deviate from each other in a smooth way, meaning that there is a drift from convergence. The reason for this is not understood, and as it may be related to the interplay between the reference metric values and the gauge conditions used in this study, understanding it will require further experimentation. In any case, the drift from 4th order is less pronounced for more accurate simulations (solid line vs. dashed and dotted lines). This suggests that convergence will be recovered better for more accurate runs, so this is not a worrying feature.

A norm convergence plot for the ``layvar'' hyperboloidal layer reference metric with and without initial perturbation is not included. It would show 1st order convergence instead of the expected 4th order. The bottom-right plot in \fref{convfig} is included to shed light on the origin of this loss of convergence, which originates at the matching between the Cauchy and the hyperboloidal layer parts of the slice. The pointwise convergence plot for the $A_{rr}$ variable shows three curves: the solid one is the difference between  medium (1200-point) and high (3600-point) resolution runs, the dotted one corresponds to the differences between low (400-point) and medium runs rescaled assuming 4th order convergence, and the dashed one is the low-med differences assuming 1st order convergence. The quantity $A_{rr}$ is chosen to be depicted due to the presence of a steep gradient in its initial data (see right plot in \fref{layvarfigs}). What seems to happen is that the numerical errors around the steep gradient next to the matching point cause the loss of convergence, and their influence is propagated to the rest of the grid during the evolution. The pointwise convergence depicted corresponds to a rather early time ($t=0.2$) to show this effect before it has propagated to the whole spatial domain. Close to the origin and \scrip the solid and dotted lines coincide, stating that convergence is still of 4th order at those regions in the grid. However, for $r\in(0.15,0.7)$ it is the dashed line that overlaps with the solid one, meaning that the convergence recovered there is only 1st order. The steepest profiles in the evolution variables happen around $r\sim 0.35$ (again, see right plot in \fref{layvarfigs}) close to the matching point between the Cauchy and hyperboloidal parts of the slice. The conclusion is that even if the matching provides continuous values for all evolution variables, it is not smooth enough to provide satisfying convergence with the current setup. 

The cases ``$c_+=1$'' and ``cos'' (not included) show especially bad errors near the origin, due to their parity not being the usual one in the used formalism, so their convergence is not expected to be good. 
Convergence could not be straightforwardly measured for the ``sim'', ``mod'', ``layCMC'' and ``laymod'' reference metrics, for the practical reason that their compactification factors were calculated using Mathematica's \texttt{NDSolve} function, which does not provide any error estimates of the solution. Implementing a different solver will allow to check convergence, but this is left for future work. 

\section{Summary and outlook}\label{concl}

In the quest for robust hyperboloidal free evolutions, having several reliable setups to choose from is a valuable asset. The present study has considered 10 different 4D hyperboloidal reference metrics of Minkowski spacetime and tested them for a specific choice of gauge conditions and with the BSSN and Z4c equations in spherical symmetry. Long-term stability has been obtained for all of them with Z4c, even if the late-time behaviour of the constraints showed relevant violations in some cases. Important properties of the background slices for the conformal compactification approach used here are the parity of the height function at the origin (for simulations with a regular center) and the smoothness of the matching point in Cauchy-hyperboloidal layer setups. The requirement $\hat\gamma_{rr}=1$, although convenient from an implementation point of view, does not seem to be a decisive factor in the long-term stability of the reference metric choices. 
This is also the first time that hyperboloidal layers have been used in non-linear simulations of the Einstein equations, giving promising results. The study of relevant aspects such as the smoothness, the location of the matching point (and thus the thickness of the layer), and the recovered convergence are left for future experiments. 
The ``CMC'' case, already used in previous work, seems to perform best in this setup as compared to the other options. However, the experiments presented are not decisive enough to suggest a clear ranking of the reference metrics considered.

One aspect of the construction that was not explored and was left for future work is the choice of different conformal and compactification factors, as all setups used $\Omega=\aconf$. There are plans to extend the study to other gauge conditions, such as those in \cite{\papbh} or the preferred-conformal-gauge-satisfying conditions in \cite{\papgauge}. Of special interest is the long-term stability behaviour when starting the simulations with Schwarzschild trumpet initial data in puncture form: are any of the reference metrics considered here advantageous? Developing non-CMC hyperboloidal trumpet data for Schwarzschild based on these reference metrics could also improve the strong field simulations. Some of the reference metrics considered here will be explored in the 3D hyperboloidal implementation, to test whether their long-term stability also holds there and to see whether they help with the convergence.

Further work will include a massless scalar field in the evolutions. It was not considered in the present work, because the current initial data only support the ``CMC'' reference metric. The initial-data solving procedure needs to be extended to consider a generic 4D reference metric to provide constraint-satisfying initial data with a scalar field perturbation. This will allow to study the scalar field's signals at \scrip, and determine whether any of the reference metrics excel in the accuracy or convergence provided there. Simulations of gravitational collapse of the scalar field will also allow to test the reference metrics during the numerically challenging dynamical part, as well as the long-term stability of the late-time solution.

\section*{Acknowledgements}

The authors thank An{\i}l Zengino\u{g}lu and David Hilditch for valuable comments on the manuscript.
AVV thanks the Fundac\~ao para a  Ci\^encia e Tecnologia (FCT), Portugal, for the financial support to the Center for Astrophysics and Gravitation (CENTRA/IST/ULisboa) through the Grant Project~No.~UIDB/00099/2020, and for the FCT funding with DOI 10.54499/DL57/2016/CP1384/CT0090. 
Acknowledged is also the financial support by the EU's H2020 ERC Advanced Grant Gravitas-101052587, and that from the VILLUM Foundation (grant VIL37766) and the DNRF Chair program (grant DNRF162) by the Danish National Research Foundation. This project received funding from EU's Horizon 2020 MSCA grants 101007855 and 101131233.

\begin{appendices}

\section{Visual comparison of height functions and reference metrics}\label{pics}

\begin{figure}[h]
\centering
\includegraphics[width=0.32\linewidth]{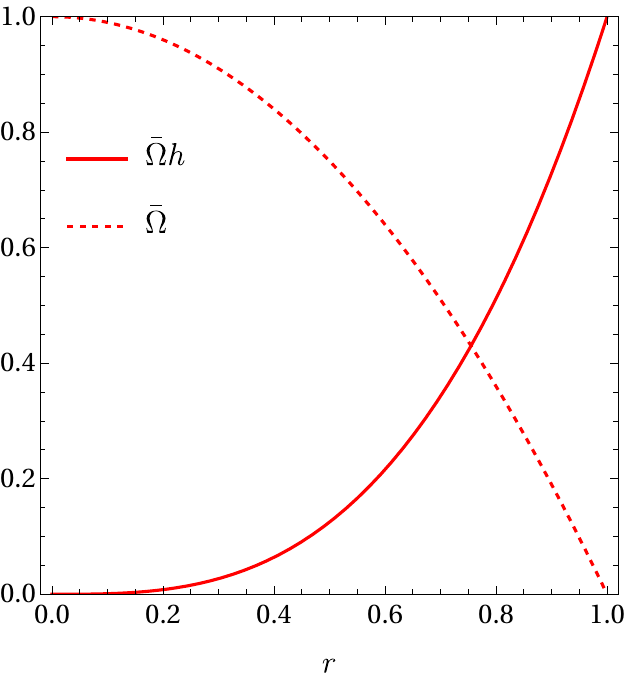}
\includegraphics[width=0.32\linewidth]{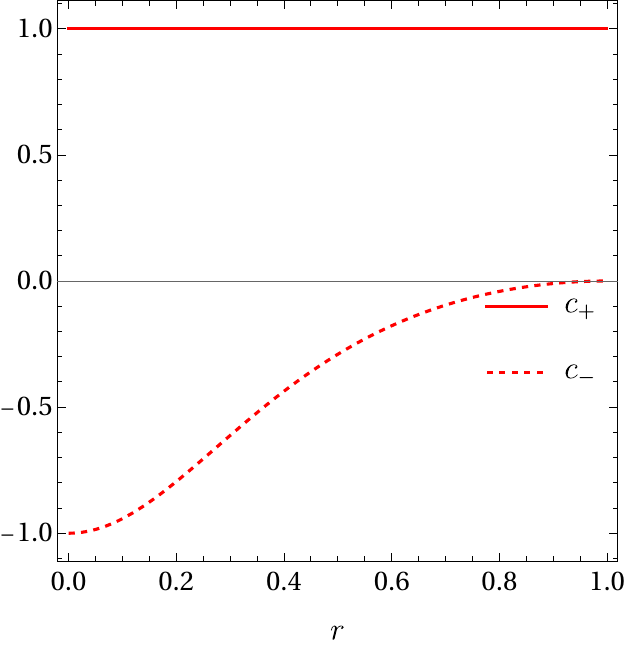}
\includegraphics[width=0.32\textwidth]{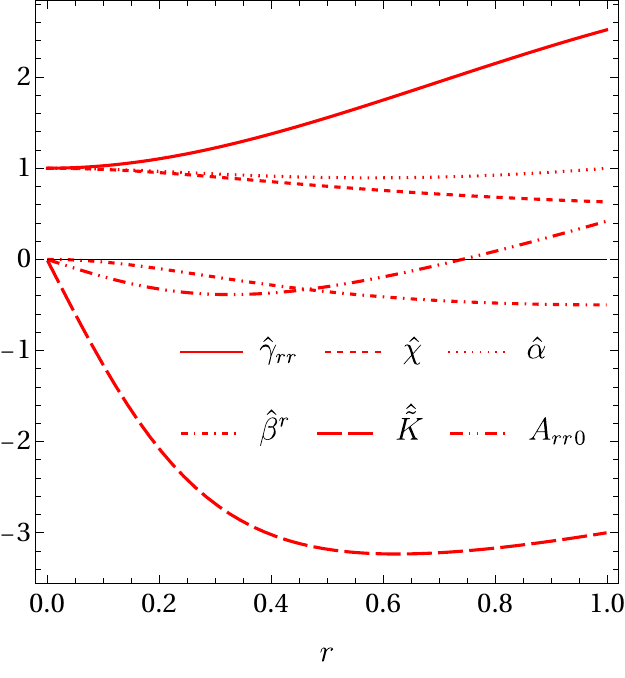}
\caption{Reference metric ``$c_+=1$'' from \ssref{cplusss}: rescaled height function $h$ \eref{cplush} and compactification factor $\aconf$ \eref{cplusaconf} on the left, physical propagation speeds $c_\pm$ \eref{speeds} on the conformally compactified hyperboloidal slice in the center, and relevant reference metric quantities and initial value for $A_{rr}$ on the right. The requirement $\hat{\gamma}_{rr}=1$ cannot be used and the height function is odd at the origin, but the outgoing propagation speed is always unity.}
\label{cplusfigs}
\vspace{2em}
\centering
\includegraphics[width=0.32\linewidth]{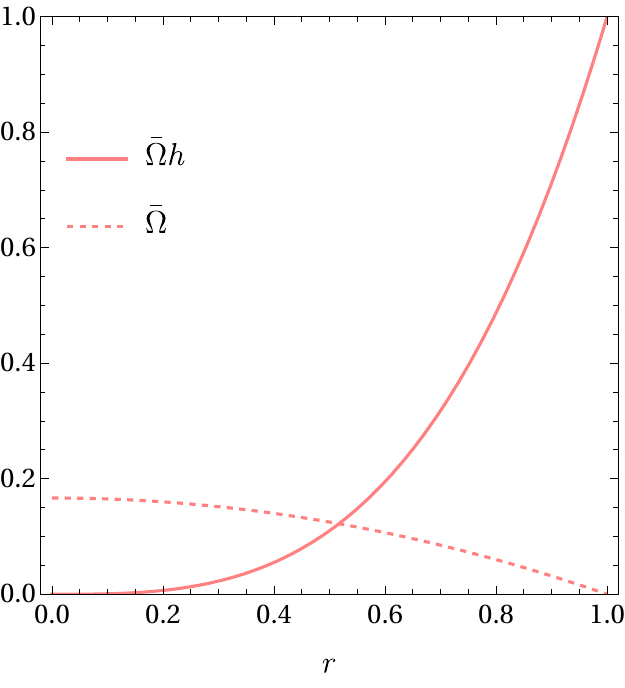}
\includegraphics[width=0.32\linewidth]{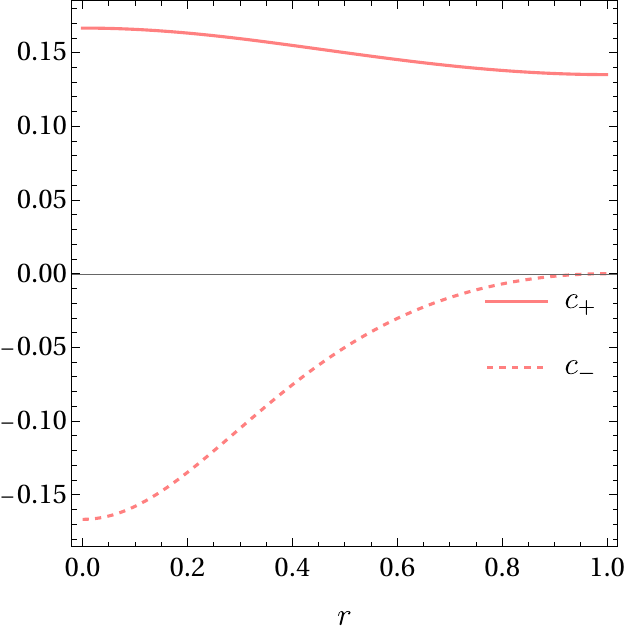}
\includegraphics[width=0.32\textwidth]{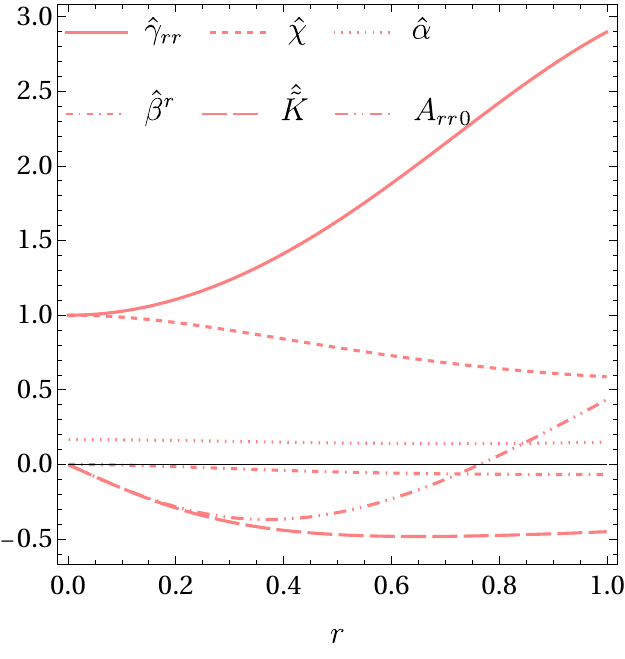}
\caption{Reference metric ``cos'' from \ssref{cossinss}: rescaled height function $h$ \eref{cosh} and compactification factor $\aconf$ \eref{cmcaconf} with $\K=-1$ on the left, physical propagation speeds $c_\pm$ on the conformally compactified hyperboloidal slice in the center, and relevant reference metric quantities and initial value for $A_{rr}$ on the right. The requirement $\hat{\gamma}_{rr}=1$ cannot be imposed and the height function is odd at the origin.}
\label{cosfigs}
\vspace{2em}
\centering
\includegraphics[width=0.32\linewidth]{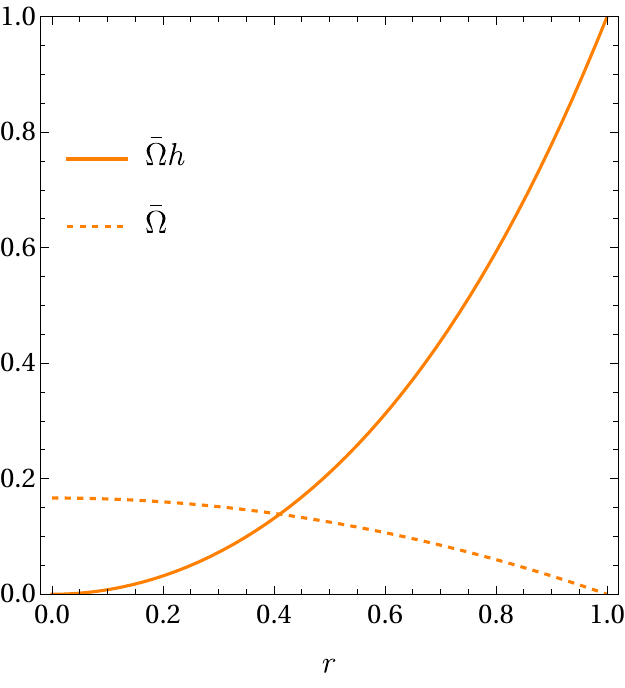}
\includegraphics[width=0.32\linewidth]{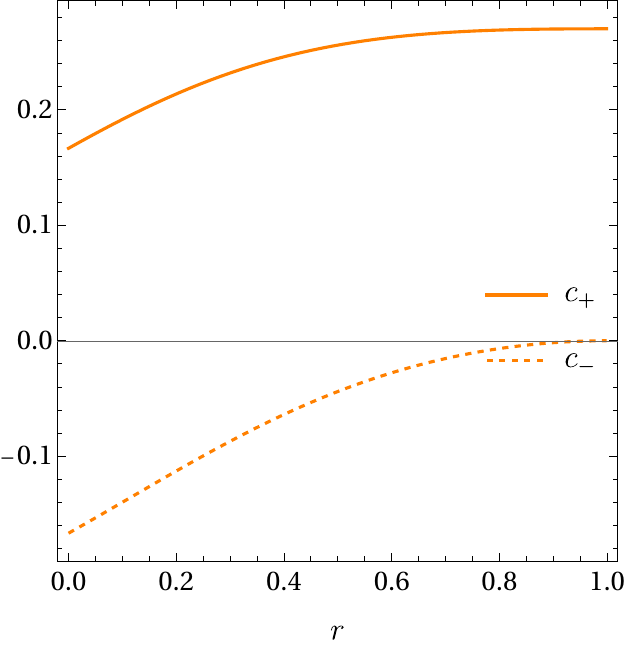}
\includegraphics[width=0.32\textwidth]{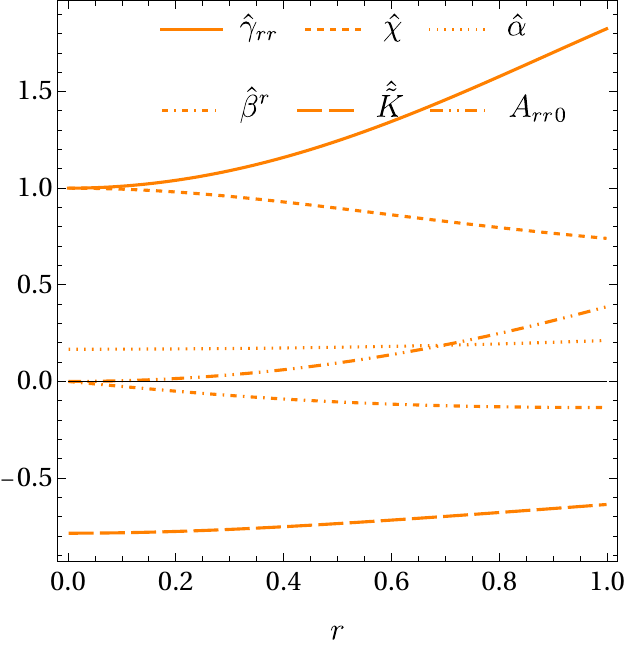}
\caption{Reference metric ``sin'' from \ssref{cossinss}: rescaled height function $h$ \eref{sinh} and compactification factor $\aconf$ \eref{cmcaconf} with $\K=-1$ on the left, physical propagation speeds $c_\pm$ on the conformally compactified hyperboloidal slice in the center, and relevant reference metric quantities and initial value for $A_{rr}$ on the right. The requirement $\hat{\gamma}_{rr}=1$ cannot be imposed.}
\label{sinfigs}
\end{figure}

\begin{figure}[h]
\centering
\includegraphics[width=0.32\linewidth]{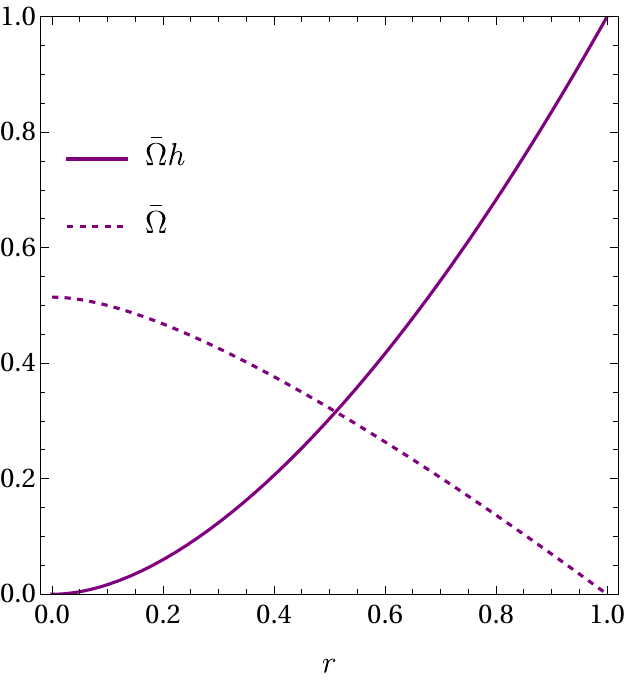}
\includegraphics[width=0.32\linewidth]{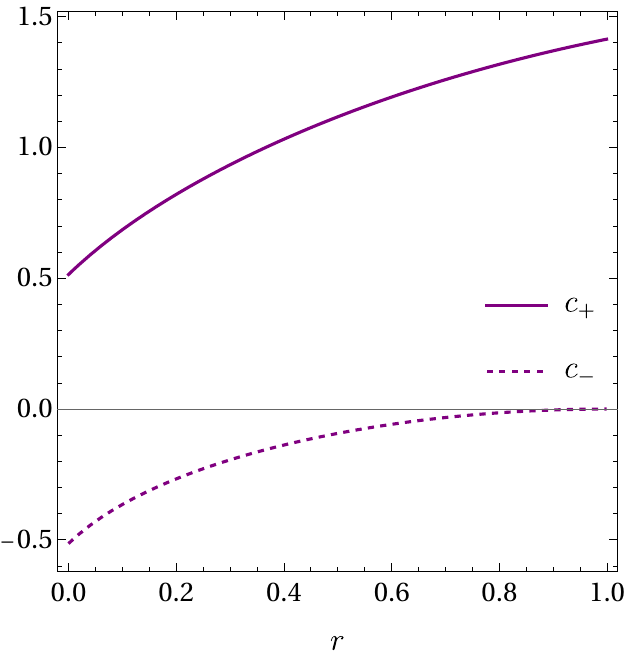}
\includegraphics[width=0.32\textwidth]{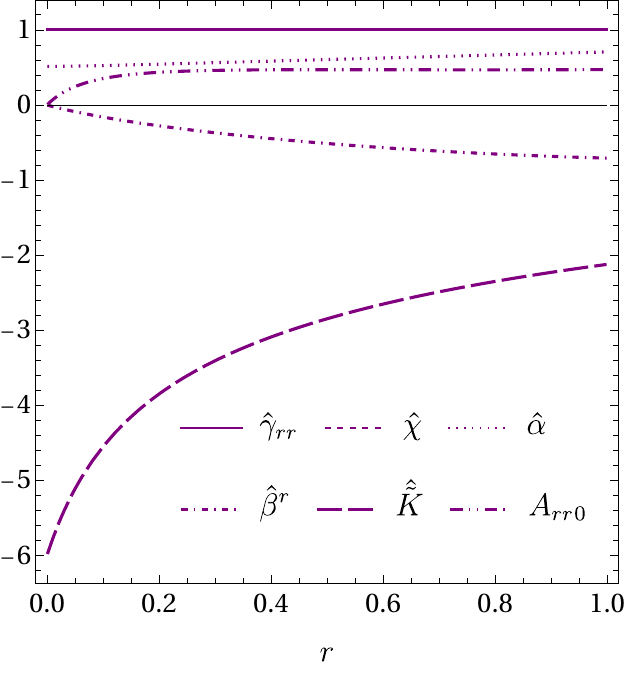}
\caption{Reference metric ``sim'' from \ssref{simss}: rescaled height function $h$ \eref{simh} with $C=1$ and compactification factor $\aconf$ satisfying $\hat{\gamma}_{rr}=1$ on the left, physical propagation speeds $c_\pm$ \eref{speeds} on the conformally compactified hyperboloidal slice in the center, and relevant reference metric quantities and initial value for $A_{rr}$ on the right. The height function has mixed parity at the origin.}
\label{simfigs}
\vspace{2em}
\centering
\includegraphics[width=0.32\linewidth]{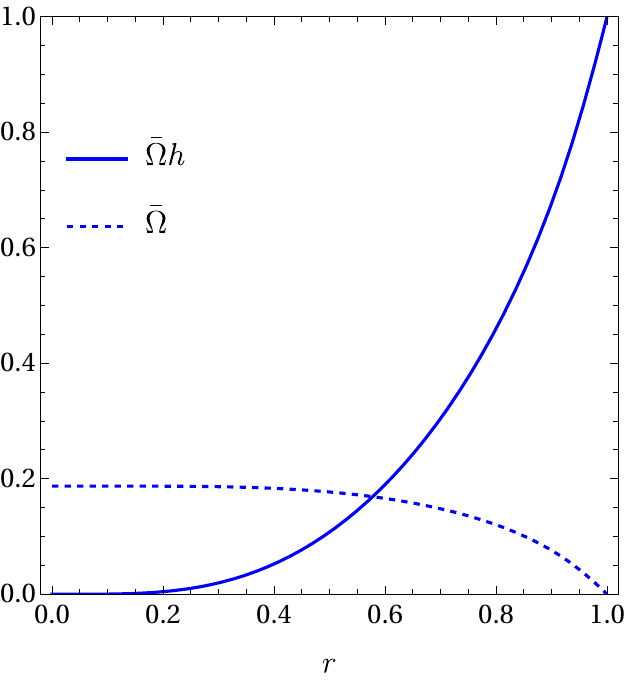}
\includegraphics[width=0.32\linewidth]{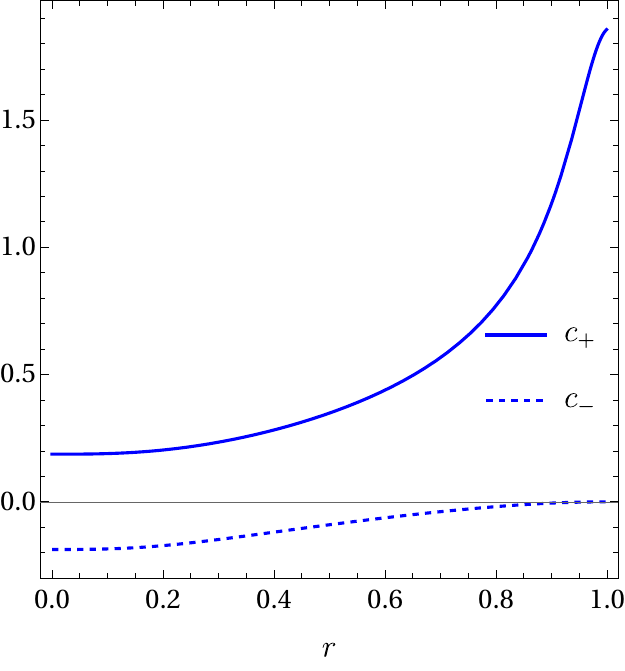}
\includegraphics[width=0.32\textwidth]{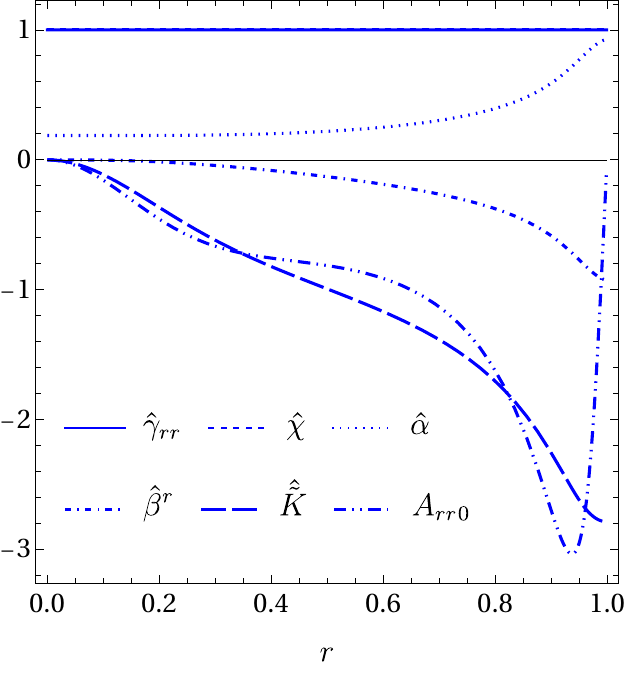}
\caption{Reference metric ``mod'' from \ssref{modss}: rescaled height function $h$ \eref{modh} with $C=2.8$ and compactification factor $\aconf$ satisfying $\hat{\gamma}_{rr}=1$ on the left, physical propagation speeds $c_\pm$ on the conformally compactified hyperboloidal slice in the center, and relevant reference metric quantities and initial value for $A_{rr}$ on the right.}
\label{modfigs}
\vspace{2em}
\centering
\includegraphics[width=0.32\linewidth]{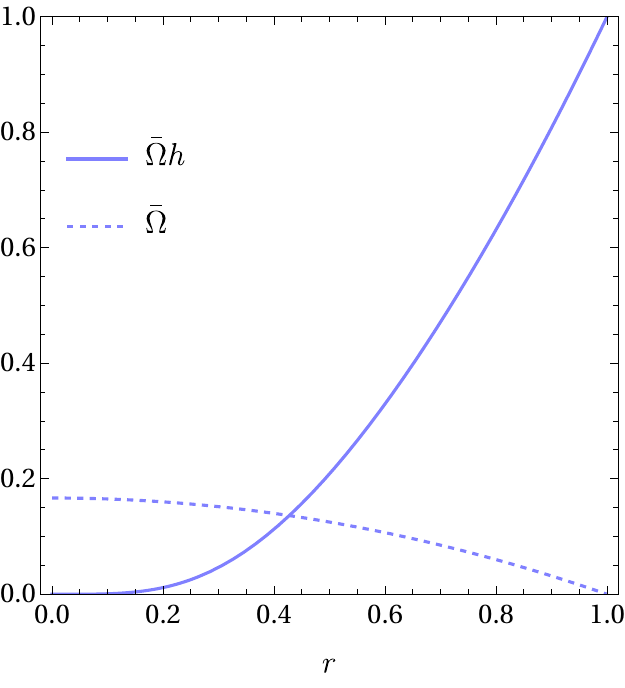}
\includegraphics[width=0.32\linewidth]{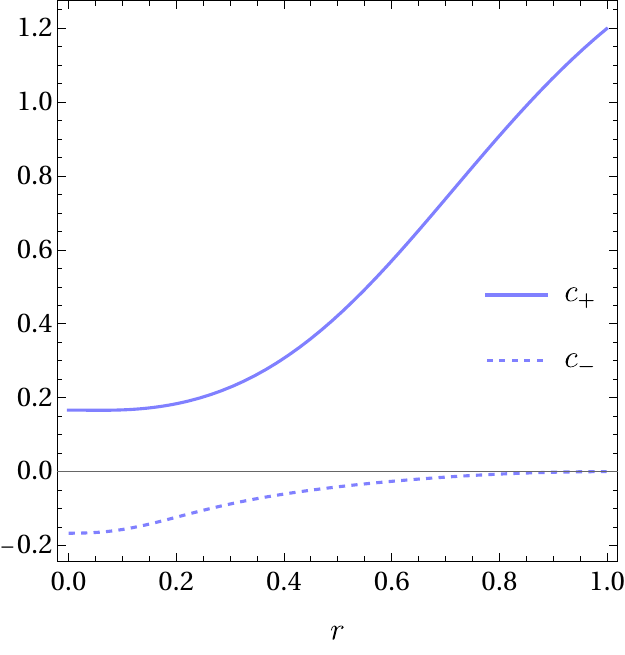}
\includegraphics[width=0.32\textwidth]{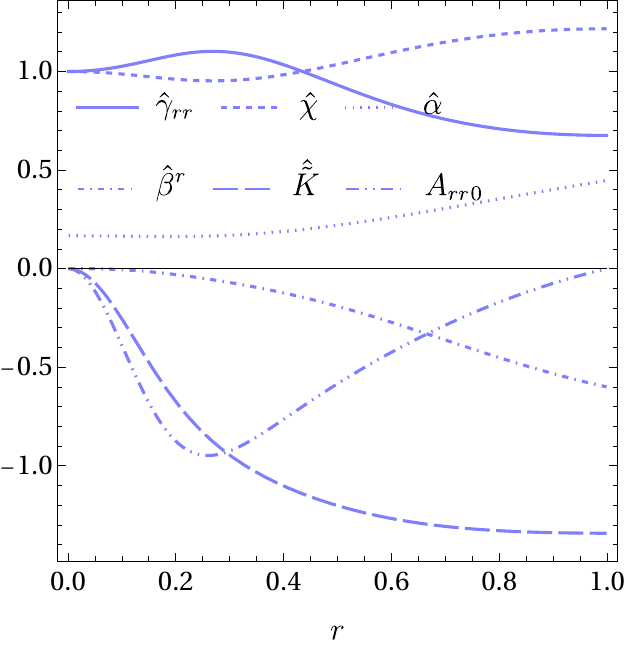}
\caption{Reference metric ``modvar'' from \ssref{modss} (variation of ``mod''): rescaled height function $h$ \eref{modh} with $C=2$ and compactification factor $\aconf$ chosen as \eref{cmcaconf} (not satisfying $\hat{\gamma}_{rr}=1$) on the left, physical propagation speeds $c_\pm$ on the conformally compactified hyperboloidal slice in the center, and relevant reference metric quantities and initial value for $A_{rr}$ on the right.}
\label{modvarfigs}
\end{figure}

\begin{figure}[h]
\centering
\includegraphics[width=0.32\linewidth]{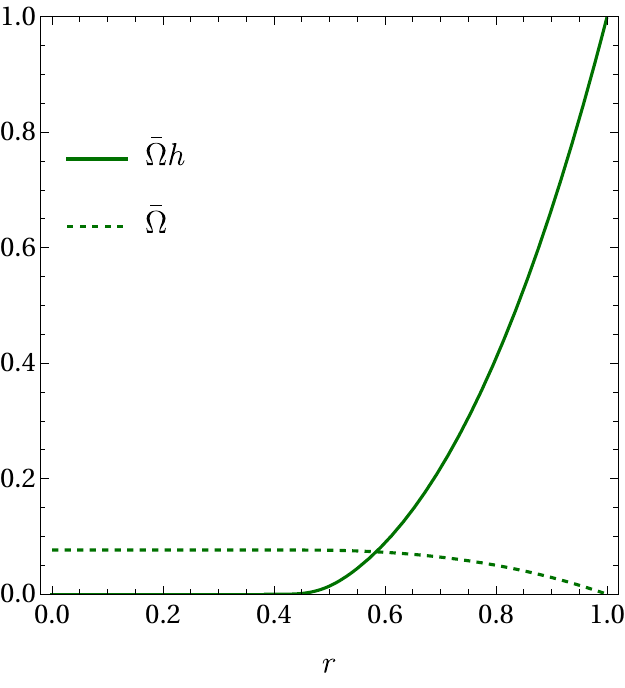}
\includegraphics[width=0.32\linewidth]{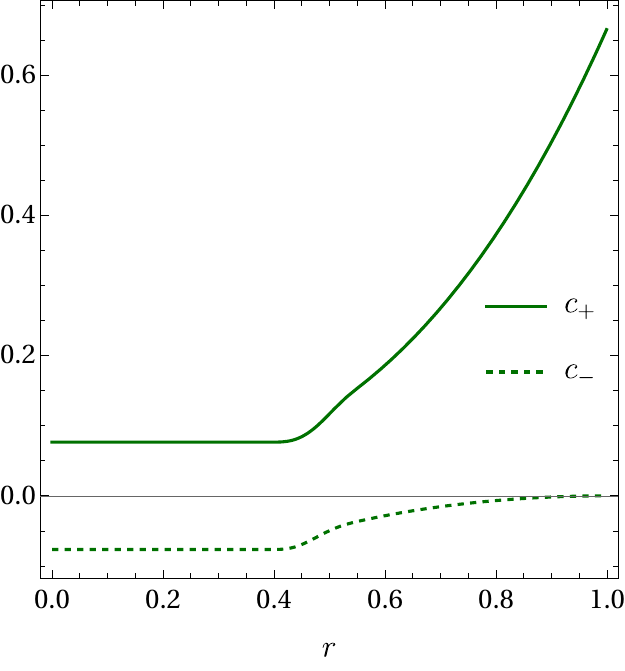}
\includegraphics[width=0.32\textwidth]{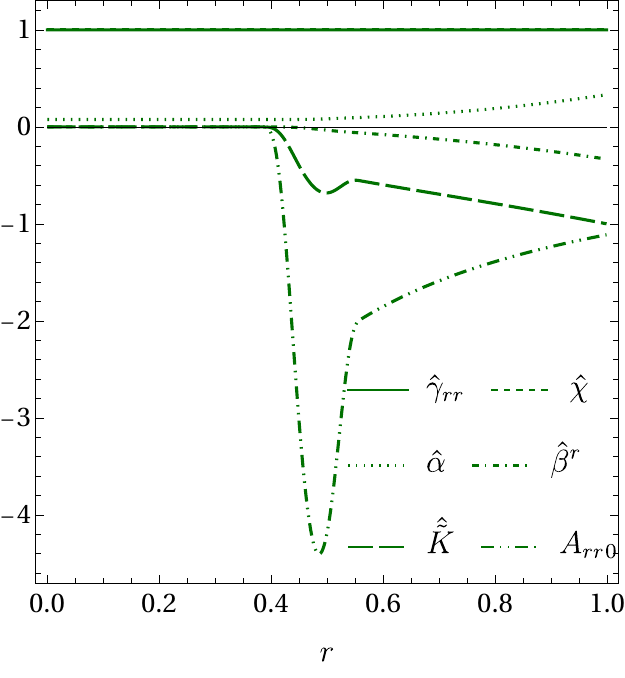}
\caption{Reference metric ``layCMC'' for hyperboloidal layer from \ssref{layalmss}: rescaled height function $h$ integrated from \eref{laysmooth} with $\rmatch=5$, $\tilde r_-=\rmatch$ and $\tilde r_+=\rmatch+2.5$, $\K=-1$ and $\rscri=1$, and corresponding numerically-solved compactification factor $\aconf$ based on \eref{layaconf} satisfying $\hat{\gamma}_{rr}=1$ on the left, physical propagation speeds $c_\pm$ \eref{speeds} on the conformally compactified hyperboloidal slice in the center, and relevant reference metric quantities and initial value for $A_{rr}$ on the right.}
\label{layalmfigs}
\vspace{2em}
\centering
\includegraphics[width=0.32\linewidth]{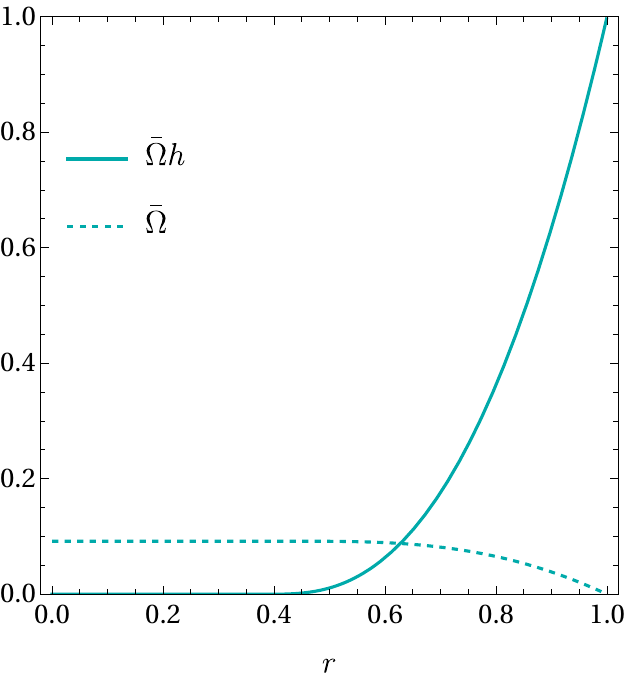}
\includegraphics[width=0.32\linewidth]{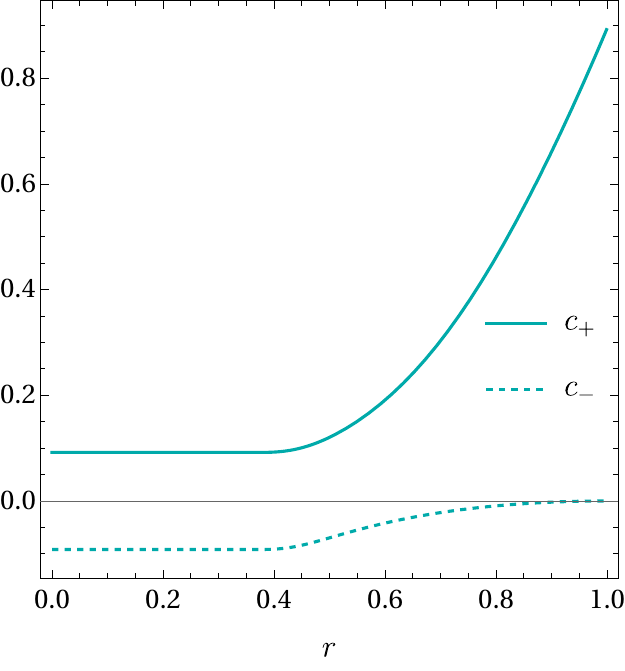}
\includegraphics[width=0.32\textwidth]{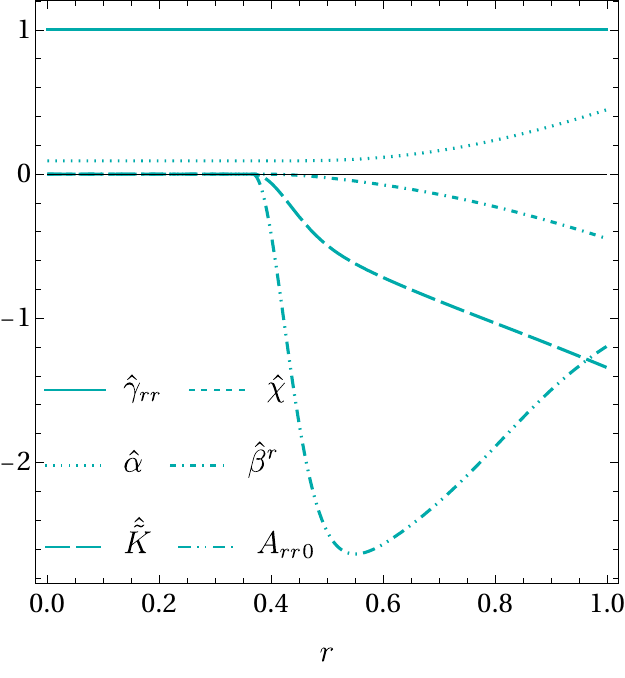}
\caption{Reference metric ``laymod'' for hyperboloidal layer from \ssref{laymodss}: rescaled height function $h$ \eref{laymodh} with $\rmatch=4$, $C=2$, $\K=-1$ and $\rscri=1$, and  corresponding numerically-solved compactification factor $\aconf$ based on \eref{layaconf} satisfying $\hat{\gamma}_{rr}=1$ on the left, physical propagation speeds $c_\pm$ on the conformally compactified hyperboloidal slice in the center, and relevant reference metric quantities and initial value for $A_{rr}$ on the right. No smoothing function needed in the matching.}
\label{laymodfigs}
\vspace{2em}
\centering
\includegraphics[width=0.32\linewidth]{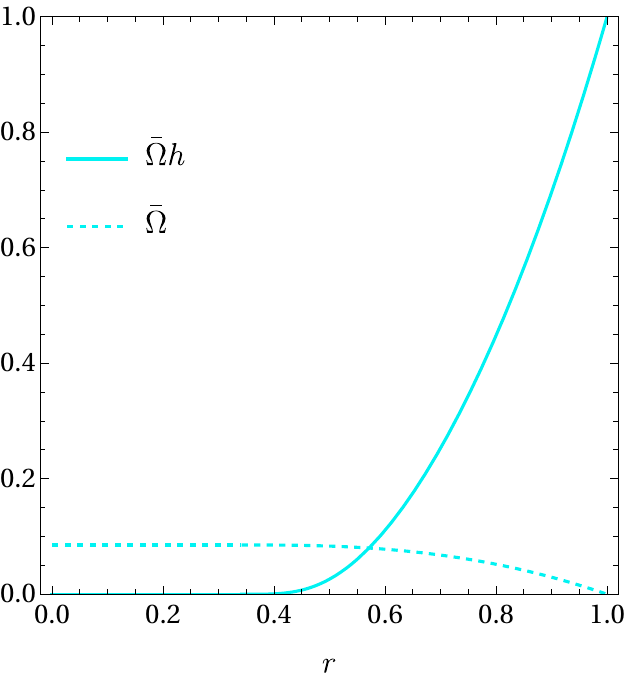}
\includegraphics[width=0.32\linewidth]{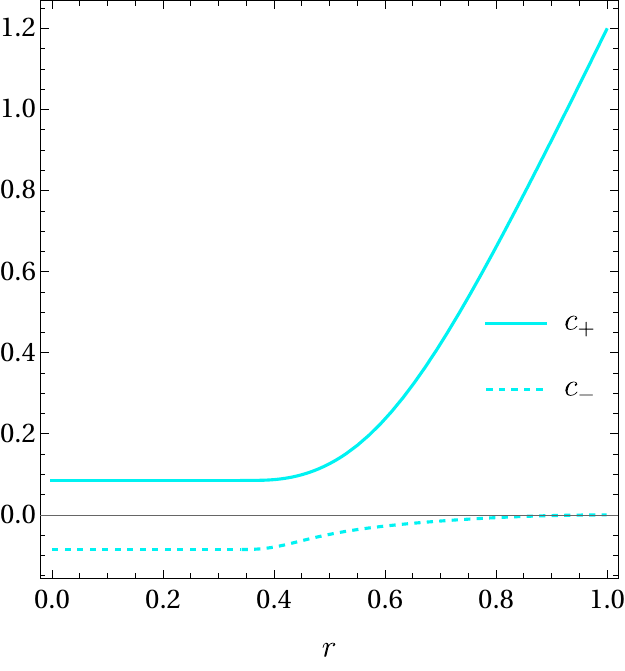}
\includegraphics[width=0.32\textwidth]{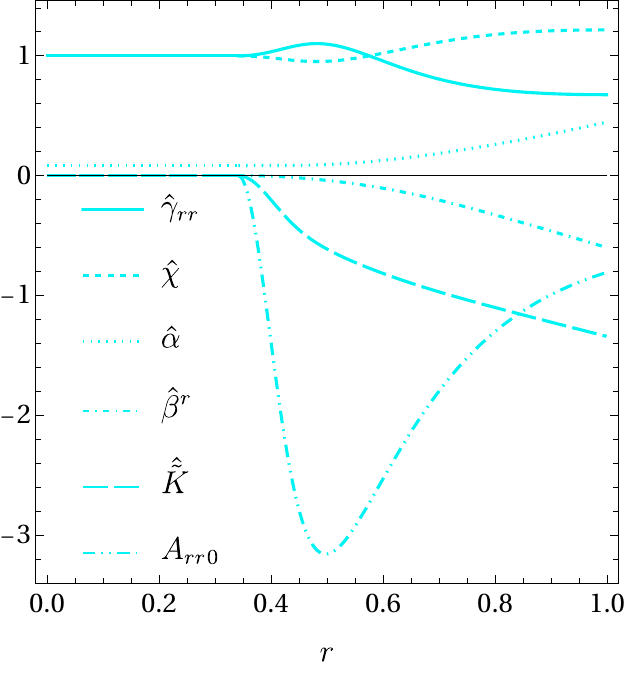}
\caption{
Reference metric ``layvar'' for hyperboloidal layer from \ssref{laymodss} (variation of ``laymod''): rescaled height function $h$ \eref{laymodh} with $\rmatch=4$, $C=2$, $\K=-1$ and $\rscri=1$, and compactification factor $\aconf$ chosen as \eref{layaconf} (not satisfying $\hat{\gamma}_{rr}=1$) on the left, physical propagation speeds $c_\pm$ on the conformally compactified hyperboloidal slice in the center, and relevant reference metric quantities and initial value for $A_{rr}$ on the right. No smoothing function needed in the matching.}
\label{layvarfigs}
\end{figure}

Figures \ref{cplusfigs}, \ref{cosfigs}, \ref{sinfigs}, \ref{simfigs}, \ref{modfigs}, \ref{modvarfigs}, \ref{layalmfigs}, \ref{laymodfigs} and \ref{layvarfigs} display the profiles of the 4D hyperboloidal reference metrics tested in this work, in an equivalent way as shown in \fref{cmcfigs} for the ``CMC'' case. 

\end{appendices}

\bibliography{extracted}

\begin{thebibliography}{71}
\expandafter\ifx\csname natexlab\endcsname\relax\def\natexlab#1{#1}\fi
\expandafter\ifx\csname bibnamefont\endcsname\relax
  \def\bibnamefont#1{#1}\fi
\expandafter\ifx\csname bibfnamefont\endcsname\relax
  \def\bibfnamefont#1{#1}\fi
\expandafter\ifx\csname citenamefont\endcsname\relax
  \def\citenamefont#1{#1}\fi
\expandafter\ifx\csname url\endcsname\relax
  \def\url#1{\texttt{#1}}\fi
\expandafter\ifx\csname urlprefix\endcsname\relax\def\urlprefix{URL }\fi
\providecommand{\bibinfo}[2]{#2}
\providecommand{\eprint}[2][]{\url{#2}}

\bibitem[{\citenamefont{Friedrich}(1983)}]{friedrich1983}
\bibinfo{author}{\bibfnamefont{H.}~\bibnamefont{Friedrich}},
  \bibinfo{journal}{Comm. Math. Phys.} \textbf{\bibinfo{volume}{91}},
  \bibinfo{pages}{445} (\bibinfo{year}{1983}),
  \urlprefix\url{http://projecteuclid.org/euclid.cmp/1103940664}.

\bibitem[{\citenamefont{Frauendiener}(2004)}]{lrr-2004-1}
\bibinfo{author}{\bibfnamefont{J.}~\bibnamefont{Frauendiener}},
  \bibinfo{journal}{Living Reviews in Relativity} \textbf{\bibinfo{volume}{7}}
  (\bibinfo{year}{2004}),
  \urlprefix\url{http://www.livingreviews.org/lrr-2004-1}.

\bibitem[{\citenamefont{Friedrich}(2002)}]{Friedrich:2003fq}
\bibinfo{author}{\bibfnamefont{H.}~\bibnamefont{Friedrich}},
  \emph{\bibinfo{title}{Conformal Einstein Evolution}}
  (\bibinfo{publisher}{Springer Berlin Heidelberg}, \bibinfo{address}{Berlin,
  Heidelberg}, \bibinfo{year}{2002}), pp. \bibinfo{pages}{1--50}, ISBN
  \bibinfo{isbn}{978-3-540-45818-0},
  \urlprefix\url{https://doi.org/10.1007/3-540-45818-2_1}.

\bibitem[{\citenamefont{Va{\~n}{\'o}-Vi{\~n}uales
  et~al.}(2015)\citenamefont{Va{\~n}{\'o}-Vi{\~n}uales, Husa, and
  Hilditch}}]{Vano-Vinuales:2014koa}
\bibinfo{author}{\bibfnamefont{A.}~\bibnamefont{Va{\~n}{\'o}-Vi{\~n}uales}},
  \bibinfo{author}{\bibfnamefont{S.}~\bibnamefont{Husa}}, \bibnamefont{and}
  \bibinfo{author}{\bibfnamefont{D.}~\bibnamefont{Hilditch}},
  \bibinfo{journal}{Class. Quant. Grav.} \textbf{\bibinfo{volume}{32}},
  \bibinfo{pages}{175010} (\bibinfo{year}{2015}).

\bibitem[{\citenamefont{Va{\~n}{\'o}-Vi{\~n}uales and
  Husa}(2015)}]{Vano-Vinuales:2014ada}
\bibinfo{author}{\bibfnamefont{A.}~\bibnamefont{Va{\~n}{\'o}-Vi{\~n}uales}}
  \bibnamefont{and} \bibinfo{author}{\bibfnamefont{S.}~\bibnamefont{Husa}},
  \bibinfo{journal}{J.Phys.Conf.Ser.} \textbf{\bibinfo{volume}{600}},
  \bibinfo{pages}{012061} (\bibinfo{year}{2015}).

\bibitem[{\citenamefont{Va{\~n}{\'o}-Vi{\~n}uales and
  Husa}(2017)}]{Vano-Vinuales:2016mbo}
\bibinfo{author}{\bibfnamefont{A.}~\bibnamefont{Va{\~n}{\'o}-Vi{\~n}uales}}
  \bibnamefont{and} \bibinfo{author}{\bibfnamefont{S.}~\bibnamefont{Husa}}, in
  \emph{\bibinfo{booktitle}{Proceedings, 14th Marcel Grossmann Meeting on
  Recent Developments in Theoretical and Experimental General Relativity,
  Astrophysics, and Relativistic Field Theories (MG14) (In 4 Volumes): Rome,
  Italy, July 12-18, 2015}} (\bibinfo{year}{2017}), vol.~\bibinfo{volume}{2},
  pp. \bibinfo{pages}{2025--2030},
  \urlprefix\url{https://inspirehep.net/record/1415732/files/arXiv:
  1601.04079.pdf}.

\bibitem[{\citenamefont{Va{\~n}{\'o}-Vi{\~n}uales and
  Husa}(2018)}]{Vano-Vinuales:2017qij}
\bibinfo{author}{\bibfnamefont{A.}~\bibnamefont{Va{\~n}{\'o}-Vi{\~n}uales}}
  \bibnamefont{and} \bibinfo{author}{\bibfnamefont{S.}~\bibnamefont{Husa}},
  \bibinfo{journal}{Class. Quant. Grav.} \textbf{\bibinfo{volume}{35}},
  \bibinfo{pages}{045014} (\bibinfo{year}{2018}).

\bibitem[{\citenamefont{Va{\~n}{\'o}-Vi{\~n}uales}(2023)}]{Vano-Vinuales:2023yzs}
\bibinfo{author}{\bibfnamefont{A.}~\bibnamefont{Va{\~n}{\'o}-Vi{\~n}uales}},
  \bibinfo{journal}{Front. Appl. Math. Stat., Sec. Statistical and
  Computational Physics} \textbf{\bibinfo{volume}{9}} (\bibinfo{year}{2023}).

\bibitem[{\citenamefont{Zengino\u{g}lu}(2007)}]{Zenginoglu:2007it}
\bibinfo{author}{\bibfnamefont{A.}~\bibnamefont{Zengino\u{g}lu}}, Ph.D. thesis,
  \bibinfo{school}{Max Planck Institute for Gravitational Physics (AEI)
  and University of Potsdam, Institute of Physics and Astronomy}
  (\bibinfo{year}{2007}).

\bibitem[{\citenamefont{Zengino\u{g}lu}(2008{\natexlab{a}})}]{Zenginoglu:2008pw}
\bibinfo{author}{\bibfnamefont{A.}~\bibnamefont{Zengino\u{g}lu}},
  \bibinfo{journal}{Class.Quant.Grav.} \textbf{\bibinfo{volume}{25}},
  \bibinfo{pages}{195025} (\bibinfo{year}{2008}{\natexlab{a}}).

\bibitem[{\citenamefont{Brandt and Br\"{u}gmann}(1997)}]{Brandt:1997tf}
\bibinfo{author}{\bibfnamefont{S.}~\bibnamefont{Brandt}} \bibnamefont{and}
  \bibinfo{author}{\bibfnamefont{B.}~\bibnamefont{Br\"{u}gmann}},
  \bibinfo{journal}{Phys.Rev.Lett.} \textbf{\bibinfo{volume}{78}},
  \bibinfo{pages}{3606} (\bibinfo{year}{1997}).

\bibitem[{\citenamefont{Campanelli et~al.}(2006)\citenamefont{Campanelli,
  Lousto, Marronetti, and Zlochower}}]{Campanelli:2005dd}
\bibinfo{author}{\bibfnamefont{M.}~\bibnamefont{Campanelli}},
  \bibinfo{author}{\bibfnamefont{C.}~\bibnamefont{Lousto}},
  \bibinfo{author}{\bibfnamefont{P.}~\bibnamefont{Marronetti}},
  \bibnamefont{and}
  \bibinfo{author}{\bibfnamefont{Y.}~\bibnamefont{Zlochower}},
  \bibinfo{journal}{Phys.Rev.Lett.} \textbf{\bibinfo{volume}{96}},
  \bibinfo{pages}{111101} (\bibinfo{year}{2006}).

\bibitem[{\citenamefont{Baker et~al.}(2006)\citenamefont{Baker, Centrella,
  Choi, Koppitz, and van Meter}}]{Baker:2005vv}
\bibinfo{author}{\bibfnamefont{J.~G.} \bibnamefont{Baker}},
  \bibinfo{author}{\bibfnamefont{J.}~\bibnamefont{Centrella}},
  \bibinfo{author}{\bibfnamefont{D.-I.} \bibnamefont{Choi}},
  \bibinfo{author}{\bibfnamefont{M.}~\bibnamefont{Koppitz}}, \bibnamefont{and}
  \bibinfo{author}{\bibfnamefont{J.}~\bibnamefont{van Meter}},
  \bibinfo{journal}{Phys.Rev.Lett.} \textbf{\bibinfo{volume}{96}},
  \bibinfo{pages}{111102} (\bibinfo{year}{2006}).

\bibitem[{\citenamefont{Hilditch}(2015)}]{Hilditch:2015qea}
\bibinfo{author}{\bibfnamefont{D.}~\bibnamefont{Hilditch}}
  (\bibinfo{year}{2015}).

\bibitem[{\citenamefont{Peterson et~al.}(2024)\citenamefont{Peterson, Gautam,
  Va{\~n}{\'o}-Vi{\~n}uales, and Hilditch}}]{Peterson:2024bxk}
\bibinfo{author}{\bibfnamefont{C.}~\bibnamefont{Peterson}},
  \bibinfo{author}{\bibfnamefont{S.}~\bibnamefont{Gautam}},
  \bibinfo{author}{\bibfnamefont{A.}~\bibnamefont{Va{\~n}{\'o}-Vi{\~n}uales}},
  \bibnamefont{and} \bibinfo{author}{\bibfnamefont{D.}~\bibnamefont{Hilditch}}
  (\bibinfo{year}{2024}).

\bibitem[{\citenamefont{Bardeen et~al.}(2011)\citenamefont{Bardeen, Sarbach,
  and Buchman}}]{Bardeen:2011ip}
\bibinfo{author}{\bibfnamefont{J.~M.} \bibnamefont{Bardeen}},
  \bibinfo{author}{\bibfnamefont{O.}~\bibnamefont{Sarbach}}, \bibnamefont{and}
  \bibinfo{author}{\bibfnamefont{L.~T.} \bibnamefont{Buchman}},
  \bibinfo{journal}{Phys.Rev.} \textbf{\bibinfo{volume}{D83}},
  \bibinfo{pages}{104045} (\bibinfo{year}{2011}).

\bibitem[{\citenamefont{Rinne and Moncrief}(2013)}]{Rinne:2013qc}
\bibinfo{author}{\bibfnamefont{O.}~\bibnamefont{Rinne}} \bibnamefont{and}
  \bibinfo{author}{\bibfnamefont{V.}~\bibnamefont{Moncrief}},
  \bibinfo{journal}{Class.Quant.Grav.} \textbf{\bibinfo{volume}{30}},
  \bibinfo{pages}{095009} (\bibinfo{year}{2013}).

\bibitem[{\citenamefont{Frauendiener et~al.}(2024)\citenamefont{Frauendiener,
  Goodenbour, and Stevens}}]{Frauendiener:2023ltp}
\bibinfo{author}{\bibfnamefont{J.}~\bibnamefont{Frauendiener}},
  \bibinfo{author}{\bibfnamefont{A.}~\bibnamefont{Goodenbour}},
  \bibnamefont{and} \bibinfo{author}{\bibfnamefont{C.}~\bibnamefont{Stevens}},
  \bibinfo{journal}{Class. Quant. Grav.} \textbf{\bibinfo{volume}{41}},
  \bibinfo{pages}{065005} (\bibinfo{year}{2024}).

\bibitem[{\citenamefont{Friedrich and Rendall}(2000)}]{Friedrich:2000qv}
\bibinfo{author}{\bibfnamefont{H.}~\bibnamefont{Friedrich}} \bibnamefont{and}
  \bibinfo{author}{\bibfnamefont{A.~D.} \bibnamefont{Rendall}},
  \emph{\bibinfo{title}{{The Cauchy Problem for the Einstein Equations}}}
  (\bibinfo{publisher}{Springer Berlin Heidelberg}, \bibinfo{address}{Berlin,
  Heidelberg}, \bibinfo{year}{2000}), vol. \bibinfo{volume}{540}, pp.
  \bibinfo{pages}{127--224}, ISBN \bibinfo{isbn}{978-3-540-46580-5},
  \urlprefix\url{https://doi.org/10.1007/3-540-46580-4_2}.

\bibitem[{\citenamefont{Bona et~al.}(1995)\citenamefont{Bona, Mass\'o, Seidel,
  and Stela}}]{Bona:1994dr}
\bibinfo{author}{\bibfnamefont{C.}~\bibnamefont{Bona}},
  \bibinfo{author}{\bibfnamefont{J.}~\bibnamefont{Mass\'o}},
  \bibinfo{author}{\bibfnamefont{E.}~\bibnamefont{Seidel}}, \bibnamefont{and}
  \bibinfo{author}{\bibfnamefont{J.}~\bibnamefont{Stela}},
  \bibinfo{journal}{Phys.Rev.Lett.} \textbf{\bibinfo{volume}{75}},
  \bibinfo{pages}{600} (\bibinfo{year}{1995}).

\bibitem[{\citenamefont{Alcubierre et~al.}(2003)\citenamefont{Alcubierre,
  Br\"{u}gmann, Diener, Koppitz, Pollney et~al.}}]{Alcubierre:2002kk}
\bibinfo{author}{\bibfnamefont{M.}~\bibnamefont{Alcubierre}},
  \bibinfo{author}{\bibfnamefont{B.}~\bibnamefont{Br\"{u}gmann}},
  \bibinfo{author}{\bibfnamefont{P.}~\bibnamefont{Diener}},
  \bibinfo{author}{\bibfnamefont{M.}~\bibnamefont{Koppitz}},
  \bibinfo{author}{\bibfnamefont{D.}~\bibnamefont{Pollney}},
  \bibnamefont{et~al.}, \bibinfo{journal}{Phys.Rev.}
  \textbf{\bibinfo{volume}{D67}}, \bibinfo{pages}{084023}
  (\bibinfo{year}{2003}).

\bibitem[{\citenamefont{Bonazzola et~al.}(2004)\citenamefont{Bonazzola,
  Gourgoulhon, Grandclement, and Novak}}]{Bonazzola:2003dm}
\bibinfo{author}{\bibfnamefont{S.}~\bibnamefont{Bonazzola}},
  \bibinfo{author}{\bibfnamefont{E.}~\bibnamefont{Gourgoulhon}},
  \bibinfo{author}{\bibfnamefont{P.}~\bibnamefont{Grandclement}},
  \bibnamefont{and} \bibinfo{author}{\bibfnamefont{J.}~\bibnamefont{Novak}},
  \bibinfo{journal}{Phys. Rev. D} \textbf{\bibinfo{volume}{70}},
  \bibinfo{pages}{104007} (\bibinfo{year}{2004}).

\bibitem[{\citenamefont{Garfinkle et~al.}(2008)\citenamefont{Garfinkle,
  Gundlach, and Hilditch}}]{Garfinkle:2007yt}
\bibinfo{author}{\bibfnamefont{D.}~\bibnamefont{Garfinkle}},
  \bibinfo{author}{\bibfnamefont{C.}~\bibnamefont{Gundlach}}, \bibnamefont{and}
  \bibinfo{author}{\bibfnamefont{D.}~\bibnamefont{Hilditch}},
  \bibinfo{journal}{Class. Quant. Grav.} \textbf{\bibinfo{volume}{25}},
  \bibinfo{pages}{075007} (\bibinfo{year}{2008}).

\bibitem[{\citenamefont{Brown}(2008)}]{Brown:2007nt}
\bibinfo{author}{\bibfnamefont{J.~D.} \bibnamefont{Brown}},
  \bibinfo{journal}{Class. Quant. Grav.} \textbf{\bibinfo{volume}{25}},
  \bibinfo{pages}{205004} (\bibinfo{year}{2008}).

\bibitem[{\citenamefont{Brown}(2009)}]{Brown:2009dd}
\bibinfo{author}{\bibfnamefont{J.~D.} \bibnamefont{Brown}},
  \bibinfo{journal}{Phys. Rev.} \textbf{\bibinfo{volume}{D79}},
  \bibinfo{pages}{104029} (\bibinfo{year}{2009}).

\bibitem[{\citenamefont{Nakamura et~al.}(1987)\citenamefont{Nakamura, Oohara,
  and Kojima}}]{NOK}
\bibinfo{author}{\bibfnamefont{T.}~\bibnamefont{Nakamura}},
  \bibinfo{author}{\bibfnamefont{K.}~\bibnamefont{Oohara}}, \bibnamefont{and}
  \bibinfo{author}{\bibfnamefont{Y.}~\bibnamefont{Kojima}},
  \bibinfo{journal}{Prog. Theor. Phys. Suppl.} \textbf{\bibinfo{volume}{90}},
  \bibinfo{pages}{1} (\bibinfo{year}{1987}).

\bibitem[{\citenamefont{Shibata and Nakamura}(1995)}]{PhysRevD.52.5428}
\bibinfo{author}{\bibfnamefont{M.}~\bibnamefont{Shibata}} \bibnamefont{and}
  \bibinfo{author}{\bibfnamefont{T.}~\bibnamefont{Nakamura}},
  \bibinfo{journal}{Phys. Rev. D} \textbf{\bibinfo{volume}{52}},
  \bibinfo{pages}{5428} (\bibinfo{year}{1995}),
  \urlprefix\url{http://link.aps.org/doi/10.1103/PhysRevD.52.5428}.

\bibitem[{\citenamefont{Baumgarte and Shapiro}(1999)}]{Baumgarte:1998te}
\bibinfo{author}{\bibfnamefont{T.~W.} \bibnamefont{Baumgarte}}
  \bibnamefont{and} \bibinfo{author}{\bibfnamefont{S.~L.}
  \bibnamefont{Shapiro}}, \bibinfo{journal}{Phys.Rev.}
  \textbf{\bibinfo{volume}{D59}}, \bibinfo{pages}{024007}
  (\bibinfo{year}{1999}).

\bibitem[{\citenamefont{Baumgarte et~al.}(2013)\citenamefont{Baumgarte,
  Montero, Cordero-Carrion, and Muller}}]{Baumgarte:2012xy}
\bibinfo{author}{\bibfnamefont{T.~W.} \bibnamefont{Baumgarte}},
  \bibinfo{author}{\bibfnamefont{P.~J.} \bibnamefont{Montero}},
  \bibinfo{author}{\bibfnamefont{I.}~\bibnamefont{Cordero-Carrion}},
  \bibnamefont{and} \bibinfo{author}{\bibfnamefont{E.}~\bibnamefont{Muller}},
  \bibinfo{journal}{Phys.Rev.} \textbf{\bibinfo{volume}{D87}},
  \bibinfo{pages}{044026} (\bibinfo{year}{2013}).

\bibitem[{\citenamefont{Montero and Cordero-Carrion}(2012)}]{Montero:2012yr}
\bibinfo{author}{\bibfnamefont{P.~J.} \bibnamefont{Montero}} \bibnamefont{and}
  \bibinfo{author}{\bibfnamefont{I.}~\bibnamefont{Cordero-Carrion}},
  \bibinfo{journal}{Phys.Rev.} \textbf{\bibinfo{volume}{D85}},
  \bibinfo{pages}{124037} (\bibinfo{year}{2012}).

\bibitem[{\citenamefont{Sanchis-Gual et~al.}(2014)\citenamefont{Sanchis-Gual,
  Montero, Font, M\"uller, and Baumgarte}}]{Sanchis-Gual:2014nha}
\bibinfo{author}{\bibfnamefont{N.}~\bibnamefont{Sanchis-Gual}},
  \bibinfo{author}{\bibfnamefont{P.~J.} \bibnamefont{Montero}},
  \bibinfo{author}{\bibfnamefont{J.~A.} \bibnamefont{Font}},
  \bibinfo{author}{\bibfnamefont{E.}~\bibnamefont{M\"uller}}, \bibnamefont{and}
  \bibinfo{author}{\bibfnamefont{T.~W.} \bibnamefont{Baumgarte}},
  \bibinfo{journal}{Phys. Rev.} \textbf{\bibinfo{volume}{D89}},
  \bibinfo{pages}{104033} (\bibinfo{year}{2014}).

\bibitem[{\citenamefont{Cordero-Carri\'on and
  Montero}(2014)}]{Cordero-Carrion:2014zza}
\bibinfo{author}{\bibfnamefont{I.}~\bibnamefont{Cordero-Carri\'on}}
  \bibnamefont{and} \bibinfo{author}{\bibfnamefont{P.~J.}
  \bibnamefont{Montero}}, \bibinfo{journal}{Springer Proc. Math. Stat.}
  \textbf{\bibinfo{volume}{60}}, \bibinfo{pages}{205} (\bibinfo{year}{2014}).

\bibitem[{\citenamefont{Montero et~al.}(2014)\citenamefont{Montero, Baumgarte,
  and M\"uller}}]{Montero:2013pca}
\bibinfo{author}{\bibfnamefont{P.~J.} \bibnamefont{Montero}},
  \bibinfo{author}{\bibfnamefont{T.~W.} \bibnamefont{Baumgarte}},
  \bibnamefont{and} \bibinfo{author}{\bibfnamefont{E.}~\bibnamefont{M\"uller}},
  \bibinfo{journal}{Phys. Rev. D} \textbf{\bibinfo{volume}{89}},
  \bibinfo{pages}{084043} (\bibinfo{year}{2014}).

\bibitem[{\citenamefont{Bona et~al.}(2003)\citenamefont{Bona, Ledvinka,
  Palenzuela, and Zacek}}]{bona-2003-67}
\bibinfo{author}{\bibfnamefont{C.}~\bibnamefont{Bona}},
  \bibinfo{author}{\bibfnamefont{T.}~\bibnamefont{Ledvinka}},
  \bibinfo{author}{\bibfnamefont{C.}~\bibnamefont{Palenzuela}},
  \bibnamefont{and} \bibinfo{author}{\bibfnamefont{M.}~\bibnamefont{Zacek}},
  \bibinfo{journal}{{Phys. Rev.}} \textbf{\bibinfo{volume}{D67}},
  \bibinfo{pages}{104005} (\bibinfo{year}{2003}),
  \urlprefix\url{doi:10.1103/PhysRevD.67.104005}.

\bibitem[{\citenamefont{Alic et~al.}(2012)\citenamefont{Alic, Bona-Casas, Bona,
  Rezzolla, and Palenzuela}}]{Alic:2011gg}
\bibinfo{author}{\bibfnamefont{D.}~\bibnamefont{Alic}},
  \bibinfo{author}{\bibfnamefont{C.}~\bibnamefont{Bona-Casas}},
  \bibinfo{author}{\bibfnamefont{C.}~\bibnamefont{Bona}},
  \bibinfo{author}{\bibfnamefont{L.}~\bibnamefont{Rezzolla}}, \bibnamefont{and}
  \bibinfo{author}{\bibfnamefont{C.}~\bibnamefont{Palenzuela}},
  \bibinfo{journal}{Phys.Rev.} \textbf{\bibinfo{volume}{D85}},
  \bibinfo{pages}{064040} (\bibinfo{year}{2012}).

\bibitem[{\citenamefont{Cao and Hilditch}(2012)}]{Cao:2011fu}
\bibinfo{author}{\bibfnamefont{Z.}~\bibnamefont{Cao}} \bibnamefont{and}
  \bibinfo{author}{\bibfnamefont{D.}~\bibnamefont{Hilditch}},
  \bibinfo{journal}{Phys. Rev. D} \textbf{\bibinfo{volume}{85}},
  \bibinfo{pages}{124032} (\bibinfo{year}{2012}).

\bibitem[{\citenamefont{DeTurck}(1983)}]{10.4310/jdg/1214509286}
\bibinfo{author}{\bibfnamefont{D.~M.} \bibnamefont{DeTurck}},
  \bibinfo{journal}{Journal of Differential Geometry}
  \textbf{\bibinfo{volume}{18}}, \bibinfo{pages}{157 } (\bibinfo{year}{1983}),
  \urlprefix\url{https://doi.org/10.4310/jdg/1214509286}.

\bibitem[{\citenamefont{Headrick et~al.}(2010)\citenamefont{Headrick, Kitchen,
  and Wiseman}}]{Headrick:2009pv}
\bibinfo{author}{\bibfnamefont{M.}~\bibnamefont{Headrick}},
  \bibinfo{author}{\bibfnamefont{S.}~\bibnamefont{Kitchen}}, \bibnamefont{and}
  \bibinfo{author}{\bibfnamefont{T.}~\bibnamefont{Wiseman}},
  \bibinfo{journal}{Class. Quant. Grav.} \textbf{\bibinfo{volume}{27}},
  \bibinfo{pages}{035002} (\bibinfo{year}{2010}).

\bibitem[{\citenamefont{Dias et~al.}(2015)\citenamefont{Dias, Santos, and
  Way}}]{Dias:2015pda}
\bibinfo{author}{\bibfnamefont{O.~J.~C.} \bibnamefont{Dias}},
  \bibinfo{author}{\bibfnamefont{J.~E.} \bibnamefont{Santos}},
  \bibnamefont{and} \bibinfo{author}{\bibfnamefont{B.}~\bibnamefont{Way}},
  \bibinfo{journal}{JHEP} \textbf{\bibinfo{volume}{04}}, \bibinfo{pages}{060}
  (\bibinfo{year}{2015}).

\bibitem[{\citenamefont{Zenginoglu}(2011)}]{Zenginoglu:2010cq}
\bibinfo{author}{\bibfnamefont{A.}~\bibnamefont{Zenginoglu}},
  \bibinfo{journal}{J. Comput. Phys.} \textbf{\bibinfo{volume}{230}},
  \bibinfo{pages}{2286} (\bibinfo{year}{2011}).

\bibitem[{\citenamefont{Bernuzzi
  et~al.}(2011{\natexlab{a}})\citenamefont{Bernuzzi, Nagar, and
  Zenginoglu}}]{Bernuzzi:2010xj}
\bibinfo{author}{\bibfnamefont{S.}~\bibnamefont{Bernuzzi}},
  \bibinfo{author}{\bibfnamefont{A.}~\bibnamefont{Nagar}}, \bibnamefont{and}
  \bibinfo{author}{\bibfnamefont{A.}~\bibnamefont{Zenginoglu}},
  \bibinfo{journal}{Phys. Rev. D} \textbf{\bibinfo{volume}{83}},
  \bibinfo{pages}{064010} (\bibinfo{year}{2011}{\natexlab{a}}).

\bibitem[{\citenamefont{Bernuzzi
  et~al.}(2011{\natexlab{b}})\citenamefont{Bernuzzi, Nagar, and
  Zengino\u{g}lu}}]{Bernuzzi:2011aj}
\bibinfo{author}{\bibfnamefont{S.}~\bibnamefont{Bernuzzi}},
  \bibinfo{author}{\bibfnamefont{A.}~\bibnamefont{Nagar}}, \bibnamefont{and}
  \bibinfo{author}{\bibfnamefont{A.}~\bibnamefont{Zengino\u{g}lu}},
  \bibinfo{journal}{Phys.Rev.} \textbf{\bibinfo{volume}{D84}},
  \bibinfo{pages}{084026} (\bibinfo{year}{2011}{\natexlab{b}}).

\bibitem[{\citenamefont{Zengino\u{g}lu and Khanna}(2011)}]{Zenginoglu:2011zz}
\bibinfo{author}{\bibfnamefont{A.}~\bibnamefont{Zengino\u{g}lu}}
  \bibnamefont{and} \bibinfo{author}{\bibfnamefont{G.}~\bibnamefont{Khanna}},
  \bibinfo{journal}{Phys.Rev.} \textbf{\bibinfo{volume}{X1}},
  \bibinfo{pages}{021017} (\bibinfo{year}{2011}).

\bibitem[{\citenamefont{Bernuzzi et~al.}(2012)\citenamefont{Bernuzzi, Nagar,
  and Zenginoglu}}]{Bernuzzi:2012ku}
\bibinfo{author}{\bibfnamefont{S.}~\bibnamefont{Bernuzzi}},
  \bibinfo{author}{\bibfnamefont{A.}~\bibnamefont{Nagar}}, \bibnamefont{and}
  \bibinfo{author}{\bibfnamefont{A.}~\bibnamefont{Zenginoglu}},
  \bibinfo{journal}{Phys. Rev. D} \textbf{\bibinfo{volume}{86}},
  \bibinfo{pages}{104038} (\bibinfo{year}{2012}).

\bibitem[{\citenamefont{Harms et~al.}(2013)\citenamefont{Harms, Bernuzzi, and
  Br\"ugmann}}]{Harms:2013ib}
\bibinfo{author}{\bibfnamefont{E.}~\bibnamefont{Harms}},
  \bibinfo{author}{\bibfnamefont{S.}~\bibnamefont{Bernuzzi}}, \bibnamefont{and}
  \bibinfo{author}{\bibfnamefont{B.}~\bibnamefont{Br\"ugmann}},
  \bibinfo{journal}{Class. Quant. Grav.} \textbf{\bibinfo{volume}{30}},
  \bibinfo{pages}{115013} (\bibinfo{year}{2013}).

\bibitem[{\citenamefont{Harms et~al.}(2014)\citenamefont{Harms, Bernuzzi,
  Nagar, and Zengino\u{g}lu}}]{Harms:2014dqa}
\bibinfo{author}{\bibfnamefont{E.}~\bibnamefont{Harms}},
  \bibinfo{author}{\bibfnamefont{S.}~\bibnamefont{Bernuzzi}},
  \bibinfo{author}{\bibfnamefont{A.}~\bibnamefont{Nagar}}, \bibnamefont{and}
  \bibinfo{author}{\bibfnamefont{A.}~\bibnamefont{Zengino\u{g}lu}},
  \bibinfo{journal}{Class.Quant.Grav.} \textbf{\bibinfo{volume}{31}},
  \bibinfo{pages}{245004} (\bibinfo{year}{2014}).

\bibitem[{\citenamefont{{Misner} et~al.}(1973)\citenamefont{{Misner}, {Thorne},
  and {Wheeler}}}]{Misner1973}
\bibinfo{author}{\bibfnamefont{C.~W.} \bibnamefont{{Misner}}},
  \bibinfo{author}{\bibfnamefont{K.~S.} \bibnamefont{{Thorne}}},
  \bibnamefont{and} \bibinfo{author}{\bibfnamefont{J.~A.}
  \bibnamefont{{Wheeler}}}, \emph{\bibinfo{title}{Gravitation}}
  (\bibinfo{publisher}{W.H.~Freeman and Co.}, \bibinfo{address}{San Francisco},
  \bibinfo{year}{1973}).

\bibitem[{\citenamefont{Penrose}(1963)}]{PhysRevLett.10.66}
\bibinfo{author}{\bibfnamefont{R.}~\bibnamefont{Penrose}},
  \bibinfo{journal}{Phys. Rev. Lett.} \textbf{\bibinfo{volume}{10}},
  \bibinfo{pages}{66} (\bibinfo{year}{1963}),
  \urlprefix\url{http://link.aps.org/doi/10.1103/PhysRevLett.10.66}.

\bibitem[{\citenamefont{Bernuzzi and Hilditch}(2010)}]{Bernuzzi:2009ex}
\bibinfo{author}{\bibfnamefont{S.}~\bibnamefont{Bernuzzi}} \bibnamefont{and}
  \bibinfo{author}{\bibfnamefont{D.}~\bibnamefont{Hilditch}},
  \bibinfo{journal}{Phys.Rev.} \textbf{\bibinfo{volume}{D81}},
  \bibinfo{pages}{084003} (\bibinfo{year}{2010}).

\bibitem[{\citenamefont{Weyhausen et~al.}(2012)\citenamefont{Weyhausen,
  Bernuzzi, and Hilditch}}]{Weyhausen:2011cg}
\bibinfo{author}{\bibfnamefont{A.}~\bibnamefont{Weyhausen}},
  \bibinfo{author}{\bibfnamefont{S.}~\bibnamefont{Bernuzzi}}, \bibnamefont{and}
  \bibinfo{author}{\bibfnamefont{D.}~\bibnamefont{Hilditch}},
  \bibinfo{journal}{Phys.Rev.} \textbf{\bibinfo{volume}{D85}},
  \bibinfo{pages}{024038} (\bibinfo{year}{2012}).

\bibitem[{\citenamefont{Va{\~n}{\'o}-Vi{\~n}uales}(2015)}]{Vano-Vinuales:2015lhj}
\bibinfo{author}{\bibfnamefont{A.}~\bibnamefont{Va{\~n}{\'o}-Vi{\~n}uales}},
  Ph.D. thesis, \bibinfo{school}{U. Illes Balears, Palma}
  (\bibinfo{year}{2015}).

\bibitem[{\citenamefont{Alcubierre}(1997)}]{Alcubierre:1996su}
\bibinfo{author}{\bibfnamefont{M.}~\bibnamefont{Alcubierre}},
  \bibinfo{journal}{Phys. Rev.} \textbf{\bibinfo{volume}{D55}},
  \bibinfo{pages}{5981} (\bibinfo{year}{1997}).

\bibitem[{\citenamefont{Baumgarte and Hilditch}(2022)}]{Baumgarte:2022ecu}
\bibinfo{author}{\bibfnamefont{T.~W.} \bibnamefont{Baumgarte}}
  \bibnamefont{and} \bibinfo{author}{\bibfnamefont{D.}~\bibnamefont{Hilditch}},
  \bibinfo{journal}{Phys. Rev. D} \textbf{\bibinfo{volume}{106}},
  \bibinfo{pages}{044014} (\bibinfo{year}{2022}).

\bibitem[{\citenamefont{Li et~al.}(2023)\citenamefont{Li, Baumgarte, Dennison,
  and de~Oliveira}}]{Li:2023pme}
\bibinfo{author}{\bibfnamefont{S.~E.} \bibnamefont{Li}},
  \bibinfo{author}{\bibfnamefont{T.~W.} \bibnamefont{Baumgarte}},
  \bibinfo{author}{\bibfnamefont{K.~A.} \bibnamefont{Dennison}},
  \bibnamefont{and} \bibinfo{author}{\bibfnamefont{H.~P.}
  \bibnamefont{de~Oliveira}}, \bibinfo{journal}{Phys. Rev. D}
  \textbf{\bibinfo{volume}{107}}, \bibinfo{pages}{064003}
  (\bibinfo{year}{2023}).

\bibitem[{\citenamefont{Gowdy}(1981)}]{10.1063/1.524975}
\bibinfo{author}{\bibfnamefont{R.~H.} \bibnamefont{Gowdy}},
  \bibinfo{journal}{Journal of Mathematical Physics}
  \textbf{\bibinfo{volume}{22}}, \bibinfo{pages}{675} (\bibinfo{year}{1981}),
  ISSN \bibinfo{issn}{0022-2488},
  \urlprefix\url{https://doi.org/10.1063/1.524975}.

\bibitem[{\citenamefont{Gentle et~al.}(2001)\citenamefont{Gentle, Holz,
  Kheyfets, Laguna, Miller et~al.}}]{Gentle:2000aq}
\bibinfo{author}{\bibfnamefont{A.~P.} \bibnamefont{Gentle}},
  \bibinfo{author}{\bibfnamefont{D.~E.} \bibnamefont{Holz}},
  \bibinfo{author}{\bibfnamefont{A.}~\bibnamefont{Kheyfets}},
  \bibinfo{author}{\bibfnamefont{P.}~\bibnamefont{Laguna}},
  \bibinfo{author}{\bibfnamefont{W.~A.} \bibnamefont{Miller}},
  \bibnamefont{et~al.}, \bibinfo{journal}{Phys.Rev.}
  \textbf{\bibinfo{volume}{D63}}, \bibinfo{pages}{064024}
  (\bibinfo{year}{2001}).

\bibitem[{\citenamefont{Malec and O'Murchadha}(2003)}]{Malec:2003dq}
\bibinfo{author}{\bibfnamefont{E.}~\bibnamefont{Malec}} \bibnamefont{and}
  \bibinfo{author}{\bibfnamefont{N.}~\bibnamefont{O'Murchadha}},
  \bibinfo{journal}{Phys.Rev.} \textbf{\bibinfo{volume}{D68}},
  \bibinfo{pages}{124019} (\bibinfo{year}{2003}).

\bibitem[{\citenamefont{Calabrese et~al.}(2006)\citenamefont{Calabrese,
  Gundlach, and Hilditch}}]{Calabrese:2005rs}
\bibinfo{author}{\bibfnamefont{G.}~\bibnamefont{Calabrese}},
  \bibinfo{author}{\bibfnamefont{C.}~\bibnamefont{Gundlach}}, \bibnamefont{and}
  \bibinfo{author}{\bibfnamefont{D.}~\bibnamefont{Hilditch}},
  \bibinfo{journal}{Class. Quant. Grav.} \textbf{\bibinfo{volume}{23}},
  \bibinfo{pages}{4829} (\bibinfo{year}{2006}).

\bibitem[{\citenamefont{Zengino\u{g}lu}(2008{\natexlab{b}})}]{Zenginoglu:2007jw}
\bibinfo{author}{\bibfnamefont{A.}~\bibnamefont{Zengino\u{g}lu}},
  \bibinfo{journal}{Class.Quant.Grav.} \textbf{\bibinfo{volume}{25}},
  \bibinfo{pages}{145002} (\bibinfo{year}{2008}{\natexlab{b}}).

\bibitem[{\citenamefont{Va{\~n}{\'o}-Vi{\~n}uales}(2024)}]{Vano-Vinuales:2023pum}
\bibinfo{author}{\bibfnamefont{A.}~\bibnamefont{Va{\~n}{\'o}-Vi{\~n}uales}},
  \bibinfo{journal}{Class. Quant. Grav.} \textbf{\bibinfo{volume}{41}},
  \bibinfo{pages}{105003} (\bibinfo{year}{2024}),
  \urlprefix\url{https://dx.doi.org/10.1088/1361-6382/ad3aca}.

\bibitem[{\citenamefont{Zengino\u{g}lu}(2024)}]{Zenginoglu:2024bzs}
\bibinfo{author}{\bibfnamefont{A.}~\bibnamefont{Zengino\u{g}lu}}
  (\bibinfo{year}{2024}).

\bibitem[{\citenamefont{Brill et~al.}(1980)\citenamefont{Brill, Cavallo, and
  Isenberg}}]{brill1980k}
\bibinfo{author}{\bibfnamefont{D.~R.} \bibnamefont{Brill}},
  \bibinfo{author}{\bibfnamefont{J.~M.} \bibnamefont{Cavallo}},
  \bibnamefont{and} \bibinfo{author}{\bibfnamefont{J.~A.}
  \bibnamefont{Isenberg}}, \bibinfo{journal}{Journal of Mathematical Physics}
  \textbf{\bibinfo{volume}{21}}, \bibinfo{pages}{2789} (\bibinfo{year}{1980}).

\bibitem[{\citenamefont{Beig and O'Murchadha}(1998)}]{Beig:1997fp}
\bibinfo{author}{\bibfnamefont{R.}~\bibnamefont{Beig}} \bibnamefont{and}
  \bibinfo{author}{\bibfnamefont{N.}~\bibnamefont{O'Murchadha}},
  \bibinfo{journal}{Phys. Rev. D} \textbf{\bibinfo{volume}{57}},
  \bibinfo{pages}{4728} (\bibinfo{year}{1998}).

\bibitem[{\citenamefont{Husa}(2003)}]{Husa:2002zc}
\bibinfo{author}{\bibfnamefont{S.}~\bibnamefont{Husa}},
  \emph{\bibinfo{title}{Numerical Relativity with the Conformal Field
  Equations}} (\bibinfo{publisher}{Springer Berlin Heidelberg},
  \bibinfo{address}{Berlin, Heidelberg}, \bibinfo{year}{2003}), vol.
  \bibinfo{volume}{617}, pp. \bibinfo{pages}{159--192}, ISBN
  \bibinfo{isbn}{978-3-540-36973-8},
  \urlprefix\url{https://doi.org/10.1007/3-540-36973-2_9}.

\bibitem[{\citenamefont{Schneemann}(2006)}]{Schneemann}
\bibinfo{author}{\bibfnamefont{C.}~\bibnamefont{Schneemann}}, Master's thesis
  (\bibinfo{year}{2006}).

\bibitem[{\citenamefont{Hilditch et~al.}(2018)\citenamefont{Hilditch, Harms,
  Bugner, Rüter, and Brügmann}}]{Hilditch:2016xzh}
\bibinfo{author}{\bibfnamefont{D.}~\bibnamefont{Hilditch}},
  \bibinfo{author}{\bibfnamefont{E.}~\bibnamefont{Harms}},
  \bibinfo{author}{\bibfnamefont{M.}~\bibnamefont{Bugner}},
  \bibinfo{author}{\bibfnamefont{H.}~\bibnamefont{Rüter}}, \bibnamefont{and}
  \bibinfo{author}{\bibfnamefont{B.}~\bibnamefont{Brügmann}},
  \bibinfo{journal}{Class. Quant. Grav.} \textbf{\bibinfo{volume}{35}},
  \bibinfo{pages}{055003} (\bibinfo{year}{2018}).

\bibitem[{\citenamefont{Gasperin et~al.}(2020)\citenamefont{Gasperin, Gautam,
  Hilditch, and Va{\~n}{\'o}-Vi{\~n}uales}}]{Gasperin:2019rjg}
\bibinfo{author}{\bibfnamefont{E.}~\bibnamefont{Gasperin}},
  \bibinfo{author}{\bibfnamefont{S.}~\bibnamefont{Gautam}},
  \bibinfo{author}{\bibfnamefont{D.}~\bibnamefont{Hilditch}}, \bibnamefont{and}
  \bibinfo{author}{\bibfnamefont{A.}~\bibnamefont{Va{\~n}{\'o}-Vi{\~n}uales}},
  \bibinfo{journal}{Class. Quant. Grav.} \textbf{\bibinfo{volume}{37}},
  \bibinfo{pages}{035006} (\bibinfo{year}{2020}).

\bibitem[{\citenamefont{Gautam et~al.}(2021)\citenamefont{Gautam,
  Va{\~n}{\'o}-Vi{\~n}uales, Hilditch, and Bose}}]{Gautam:2021ilg}
\bibinfo{author}{\bibfnamefont{S.}~\bibnamefont{Gautam}},
  \bibinfo{author}{\bibfnamefont{A.}~\bibnamefont{Va{\~n}{\'o}-Vi{\~n}uales}},
  \bibinfo{author}{\bibfnamefont{D.}~\bibnamefont{Hilditch}}, \bibnamefont{and}
  \bibinfo{author}{\bibfnamefont{S.}~\bibnamefont{Bose}},
  \bibinfo{journal}{Phys. Rev. D} \textbf{\bibinfo{volume}{103}},
  \bibinfo{pages}{084045} (\bibinfo{year}{2021}).

\bibitem[{\citenamefont{Peterson et~al.}(2023)\citenamefont{Peterson, Gautam,
  Rainho, Va{\~n}{\'o}-Vi{\~n}uales, and Hilditch}}]{Peterson:2023bha}
\bibinfo{author}{\bibfnamefont{C.}~\bibnamefont{Peterson}},
  \bibinfo{author}{\bibfnamefont{S.}~\bibnamefont{Gautam}},
  \bibinfo{author}{\bibfnamefont{I.}~\bibnamefont{Rainho}},
  \bibinfo{author}{\bibfnamefont{A.}~\bibnamefont{Va{\~n}{\'o}-Vi{\~n}uales}},
  \bibnamefont{and} \bibinfo{author}{\bibfnamefont{D.}~\bibnamefont{Hilditch}},
  \bibinfo{journal}{Phys. Rev. D} \textbf{\bibinfo{volume}{108}},
  \bibinfo{pages}{024067} (\bibinfo{year}{2023}).

\bibitem[{\citenamefont{Kreiss and Oliger}(1973)}]{kreiss1973methods}
\bibinfo{author}{\bibfnamefont{H.}~\bibnamefont{Kreiss}} \bibnamefont{and}
  \bibinfo{author}{\bibfnamefont{J.}~\bibnamefont{Oliger}},
  \emph{\bibinfo{title}{Methods for the approximate solution of time dependent
  problems}}, GARP publications series No. 10
  (\bibinfo{publisher}{International Council of Scientific Unions, World
  Meteorological Organization}, \bibinfo{year}{1973}),
  \urlprefix\url{http://books.google.es/books?id=OxMZAQAAIAAJ}.

\bibitem[{\citenamefont{Babiuc et~al.}(2008)\citenamefont{Babiuc, Husa, Alic,
  Hinder, Lechner, Schnetter, Szilágyi, Zlochower, Dorband, Pollney
  et~al.}}]{Babiuc_2008}
\bibinfo{author}{\bibfnamefont{M.~C.} \bibnamefont{Babiuc}},
  \bibinfo{author}{\bibfnamefont{S.}~\bibnamefont{Husa}},
  \bibinfo{author}{\bibfnamefont{D.}~\bibnamefont{Alic}},
  \bibinfo{author}{\bibfnamefont{I.}~\bibnamefont{Hinder}},
  \bibinfo{author}{\bibfnamefont{C.}~\bibnamefont{Lechner}},
  \bibinfo{author}{\bibfnamefont{E.}~\bibnamefont{Schnetter}},
  \bibinfo{author}{\bibfnamefont{B.}~\bibnamefont{Szilágyi}},
  \bibinfo{author}{\bibfnamefont{Y.}~\bibnamefont{Zlochower}},
  \bibinfo{author}{\bibfnamefont{N.}~\bibnamefont{Dorband}},
  \bibinfo{author}{\bibfnamefont{D.}~\bibnamefont{Pollney}},
  \bibnamefont{et~al.}, \bibinfo{journal}{Classical and Quantum Gravity}
  \textbf{\bibinfo{volume}{25}}, \bibinfo{pages}{125012}
  (\bibinfo{year}{2008}),
  \urlprefix\url{https://dx.doi.org/10.1088/0264-9381/25/12/125012}.

\end{thebibliography}

\end{document}